\documentclass[12pt,a4paper]{article}
\usepackage{amsmath,amsfonts}
\usepackage{amsthm,amssymb}
\usepackage{epsfig,amsopn}
\usepackage{psfrag}
\usepackage[retainorgcmds]{IEEEtrantools}

\def\I0{\textnormal{I}_0}
\def\K{\textnormal{K}}
\def\Exp{\textnormal{Exp}}

\def\X{\mathcal{X}}
\def\Y{\mathcal{Y}}
\newcommand{\Es}{{\mathcal E}}

\newcommand{\bfnu}{\boldsymbol{\nu}}

\newcommand{\bfw}{{\vect{w}}}

\newcommand{\lam}{\boldsymbol{\lambda}}
\newcommand{\bfef}{\vect{f}}

\newcommand{\bfone}{\bf{1}}

\newcommand{\bfq}{\vect{q}}
\newcommand{\bfzero}{\vect{0}}
\newcommand{\vect}[1]{\mathbf{#1}}
\newcommand{\const}[1]{\textnormal{\usefont{T1}{bc9}{r}{r}\selectfont #1}}
\renewcommand{\d}{\,\textnormal{d}}
\newcommand{\set}[1]{\mathcal{#1}}
\def\Reals{\mathbb R}
\def\Complex{\mathbb C}
\newcommand{\E}[2][]{\textnormal{\textsf{E}}_{#1}\!\left[{#2}\right]} 
\newtheorem{proposition}{Proposition}

\newcommand{\law}[1]{\mathsf{#1}} 
\newcommand{\GalErq}{E_{\textnormal{G}}}
\newcommand{\GalEr}{E_{\textnormal{G,r}}}
\newcommand{\GalEzrhoq}{E_{\textnormal{G},0}}
\newcommand{\CKErq}{E_{\textnormal{CK}}}
\newcommand{\CKEr}{E_{\textnormal{CK,r}}}
\newcommand{\CKEzrhoq}{E_{\textnormal{CK},0}}
\newcommand{\Er}{E_{\textnormal{r}}}
\newcommand{\Ezrho}{E_{0}}
\newcommand{\Esp}{E_{\textnormal{sp}}}
\newcommand{\GalErqM}{E_{\textnormal{G,0}}^{\textnormal{M}}}
\newcommand{\GalErM}{E_{\textnormal{G,r}}^{\textnormal{M}}}
\newcommand{\eps}{\epsilon}

\newcommand{\dabs}{|d|}

\date{}

\begin{document}

\bibliographystyle{ieeetr}

\title{Duality Bounds on the Cut-Off Rate with Applications to Ricean
  Fading}

\author{Amos Lapidoth \and Natalia Miliou}

\maketitle
\begin{abstract}
 
  We propose a technique to derive upper bounds on Gallager's
  cost-constrained random coding exponent function. Applying this
  technique to the non-coherent peak-power or average-power limited
  discrete time memoryless Ricean fading channel, we obtain the high
  signal-to-noise ratio (SNR) expansion of this channel's cut-off
  rate. At high SNR the gap between channel capacity and the cut-off
  rate approaches a finite limit.  This limit is approximately 0.26
  nats per channel-use for zero specular component (Rayleigh) fading
  and approaches 0.39 nats per channel-use for very large specular
  components.
  
  We also compute the asymptotic cut-off rate of a Rayleigh fading
  channel when the receiver has access to some partial side
  information concerning the fading. It is demonstrated that the
  cut-off rate does not utilize the side information as efficiently as
  capacity, and that the high SNR gap between the two increases to
  infinity as the imperfect side information becomes more and more
  precise.
\end{abstract}


\section{Introduction}
\label{sec:intro}

This paper addresses the computation of a function that is key to the
evaluation of both the random coding and sphere packing error exponents.
This function, often denoted $\Ezrho(\varrho)$, is usually expressed
as a maximization problem over input distributions.  Consequently, it
is conceptually easily bounded from below: any feasible input
distribution gives rise to such a bound. In this paper we propose to
use a dual expression for $\Ezrho(\varrho)$ --- an expression that
involves a minimization over output distributions --- in order to
derive \emph{upper} bounds on $\Ezrho(\varrho)$. We shall demonstrate
this approach by studying the cutoff rate of non-coherent Ricean
fading channels. To that end we shall have to study the appropriate
modifications to the function $\Ezrho(\varrho)$ that are needed to
account for input constraints and when the channel input and output
alphabets are infinite.

It should be noted that the dual expression we propose to use is not
new \cite{Blahut74}, \cite[Ex.\ 23 in Ch. 2.5]{CK}. We merely extend
it here to input constrained channels over infinite alphabets and
demonstrate how it can be used to derive \emph{analytic} upper bounds
on the random coding and sphere packing error exponents. For
\emph{numerical} procedures (for unconstrained finite alphabet
channels) see \cite{Arimoto76}.

The rest of this introductory section is dedicated to the introduction
of the function $\Ezrho(\varrho)$ for discrete memoryless channels. We
first treat unconstrained channel and then introduce the modifications
that are needed to account for input constraints. We describe both the
``method of types'' approach and Gallager's approach. We pay special
attention to the modification that Gallager introduced to account for
cost constraints and to the duality between the expressions derived
using the two approaches. This introduction is somewhat lengthy
because, while the results are not new, we had difficulty pointing to
a publication that introduces the two approaches side by side and that
compares the two in the presence of cost constraints. 

In Section~\ref{sec:continuous} we extend the discussion to infinite
alphabets and prove the basic inequality on which our approach to
upper bounding $\Ezrho(\varrho)$ is based; see
Proposition~\ref{prop:upper}. In Section~\ref{sec:Ricean} we introduce
the discrete-time memoryless Ricean fading channel with and without
full or partial side information at the receiver, and we describe our
asymptotic results on this channel's cutoff rate. These asymptotic
results are derived using duality in Section~\ref{sec:derivation},
which concludes the paper.

\subsection{Unconstrained Inputs}

To motivate the interest in the function $\Ezrho(\varrho)$ we shall
begin by addressing the case where there are no input constraints.
The reliability function $E(R)$ corresponding to rate-$R$
unconstrained communication over a discrete memoryless channel (DMC)
of capacity $C \geq R$ is the best exponential decay in the blocklength $n$ of the
average probability of error that one can achieve using rate-$R$
blocklength-$n$ codebooks.  That is,
\begin{equation}
  \label{eq:amos1}
  E(R) \triangleq \varlimsup_{n \rightarrow \infty} - \frac{1}{n} \log
  \text{P}_{\text{e}}(n, R)
\end{equation}
where $\text{P}_{\text{e}}(n, R)$ denotes the average probability of error
of the best rate-$R$ blocklength-$n$ codebook for the given channel.

The problem of computing the reliability function of a general DMC
over the finite input and output alphabets $\X$ and $\Y$ and of a
general law $\law{W}(y|x)$ is still open. Various upper and lower
bounds are, however, known. To derive lower bounds on the reliability
function one must derive upper bounds on the probability of error of
the best rate-$R$ blocklength-$n$ code. This is typically done by
demonstrating the existence of good codes for which the average
probability of error is small. One such lower bound on $E(R)$ is the
random coding lower bound \cite{Gallager68}. By considering an ensemble
of codebooks whose codewords are chosen independently, each according
to a product distribution of marginal law $\law{Q}$, Gallager derived
the lower bound
\begin{equation}
  \label{eq:amos10}
  E(R) \geq \GalErq(R,\law{Q})
\end{equation}
where
\begin{equation}
  \label{eq:amos20}
  \GalErq(R,\law{Q}) \triangleq \max_{0 \leq \varrho \leq 1} \{
  \GalEzrhoq(\varrho, \law{Q}) - \varrho R\}
\end{equation}
and
\begin{equation}
  \label{eq:amos25}
  \GalEzrhoq(\varrho, \law{Q}) \triangleq - \log \sum_{y \in \Y} 
  \left(\sum_{x\in\X}
  \law{Q}(x) \law{W}(y|x)^\frac{1}{1+\varrho}\right)^{1+\varrho}.
\end{equation}
Since the law $\law{Q}$ from which the ensemble of codebooks is
constructed is arbitrary, Gallager obtained the bound
\begin{equation}
  \label{eq:amos30}
  E(R) \geq \GalEr(R)
\end{equation}
where $\GalEr(R)$ is Gallager's random coding error exponent
\begin{align}
  \GalEr(R) & \triangleq \max_{\law{Q}} \GalErq(R,\law{Q}) \\
            & = \max_{\law{Q}} \max_{0 \leq \varrho \leq 1} 
            \{ \GalEzrhoq(\varrho,\law{Q}) - \varrho R\}.
\end{align}

A different random coding lower bound on the reliability function can
be derived using the ensemble of codebooks where the codewords are
still chosen independently, but rather than according to a product
distribution, each is now chosen uniformly over a type class \cite[2.5]{CK},
\cite{Blahut74}, \cite{Poltyrev82}. With
this approach 
one obtains \cite[2.5]{CK}, \cite{Blahut74} the lower
bound
\begin{equation}
  \label{eq:amos40}
  E(R) \geq \CKErq(R,\law{Q})
\end{equation}
where 
\begin{equation}
  \label{eq:amos50}
  \CKErq(R,\law{Q}) \triangleq \min_{\law{V}(\cdot|\cdot)} \left\{ 
    D(\law{V} \| \law{W} | \law{Q}) + |I(\law{Q}, \law{V}) - R|^{+} \right\}. 
\end{equation}
Here the minimization is over all conditional laws 
\begin{equation}
 \law{V}(y|x)
\geq 0, \qquad \sum_{y \in \Y} \law{V}(y|x) = 1, \forall x \in \X; 
\end{equation}
\begin{align}
  D(\law{V} \| \law{W} | \law{Q}) & = \sum_{x \in \X} \law{Q}(x) D \bigl(
  \law{V}(\cdot|x) \| \law{W}(\cdot|x) \bigr) \\
  & = \sum_{x \in \X} \law{Q}(x) 
  \sum_{y \in \Y} \law{V}(y|x) \log \frac{\law{V}(y|x)}{\law{W}(y|x)};
\end{align}
the term $I(\law{Q}, \law{V})$ denotes the mutual information
corresponding to the channel $\law{V}$ and the input distribution
$\law{Q}$; and $|\xi|^{+}$ stands for $\max\{\xi, 0\}$. Again, since
the type $\law{Q}$ according to which the ensemble is
generated is arbitrary, one obtains
\begin{equation}
  E(R) \geq \CKEr(R)
\end{equation}
where
\begin{align}
\CKEr(R) & \triangleq \max_{\law{Q}} \CKErq(R,\law{Q}) \\
      & = \max_{\law{Q}} \min_{\law{V}(\cdot|\cdot)} \left\{ 
    D(\law{V} \| \law{W} | \law{Q}) + |I(\law{Q}, \law{V}) - R|^{+} \right\}.
\end{align}

There is an alternative form for $\CKErq(R, \law{Q})$ that will be of
interest to us \cite{Blahut74}, \cite[Ex.\ 23 in Ch.
2.5]{CK}. This form is more similar to \eqref{eq:amos20}:
\begin{equation}
  \label{eq:amos100}
  \CKErq(R,\law{Q}) = \max_{0 \leq \varrho \leq 1} \{
  \CKEzrhoq(\varrho, \law{Q}) - \varrho R\}
\end{equation}
where
\begin{align}
  \label{eq:amos102}
  \CKEzrhoq(\varrho, \law{Q}) & \triangleq \min_{\law{V}(\cdot|\cdot)} \left\{ 
    D(\law{V} \| \law{W} | \law{Q}) + \varrho I(\law{Q}, \law{V})
    \right\} \\
    \label{eq:amos104}
    & = \min_{\law{R}} \left\{ -(1+\varrho) \sum_{x \in \X} \law{Q}(x) \log \left( \sum_{y \in \Y}
  \law{W}(y|x)^{\frac{1}{1+\varrho}}
    \law{R}(y)^{\frac{\varrho}{1+\varrho}} \right) \right\}
\end{align}
and where the minimization in the latter is over the set of all
distributions $\law{R}$ on the output alphabet $\Y$.

In general, for any DMC $\law{W}(y|x)$ and any input distribution
$\law{Q}$ \cite{Blahut74}, \cite[Ex.\ 23 in Ch.  2.5]{CK}
\begin{equation}
  \label{eq:jazz10}
  \CKEzrhoq(\varrho, \law{Q}) \geq \GalEzrhoq(\varrho, \law{Q}),
  \qquad \varrho \geq 0
\end{equation}
and hence
\begin{equation}
  \label{eq:jazz20}
  \CKErq(R, \law{Q}) \geq \GalErq(R, \law{Q})
\end{equation}
with the inequalities typically being strict. These inequalities are a
consequence of the fact that the ``average constant composition code''
performs better than the ``average independent and identically
distributed code'' \cite{Gallager73}. However, when optimized over the
input distributions, the inequalities turn into equalities
\cite{Blahut74}, \cite{Csiszar98}, \cite[Ex.\ 23 in Ch.  2.5]{CK}
\begin{equation}
  \label{eq:jazz30}
  \max_{\law{Q}} \CKEzrhoq(\varrho, \law{Q}) =  \max_{\law{Q}} \GalEzrhoq(\varrho, \law{Q}),
  \qquad \varrho \geq 0
\end{equation}
and
\begin{equation}
  \label{eq:jazz40}
    \max_{\law{Q}} \CKErq(R, \law{Q}) = \max_{\law{Q}} \GalErq(R, \law{Q})
\end{equation}
i.e.,
\begin{equation}
  \label{eq:jazz50}
  \CKEr(R) = \GalEr(R).
\end{equation}
In fact, as shown in Appendix~\ref{app:lagrange}, the optimization
problems appearing on the LHS and on the RHS of \eqref{eq:jazz40} are
Lagrange duals.

Consequently, we shall henceforth denote $\max_{\law{Q}}
\CKEzrhoq(\varrho, \law{Q})$ ($= \max_{\law{Q}} \GalEzrhoq(\varrho,
\law{Q})$) by $\Ezrho(\varrho)$ and refer to $\GalEr(R)$ ($=\CKEr(R)$)
as the random coding error exponent and denote it by $\Er(R)$.  In
terms of the function $\Ezrho(\cdot)$ the random coding error exponent
$\Er(R)$ is thus given by
\begin{equation}
  \label{eq:jazz60}
  \Er(R) = \max_{0 \leq \varrho \leq 1} \{ \Ezrho(\varrho) - \varrho R\}.
\end{equation}

The  cut-off rate $R_{0}$ is defined by 
\begin{equation}
  \label{eq:def_R0}
  R_{0} = \Bigl. \Ezrho(\varrho) \Bigr|_{\varrho = 1}.
\end{equation}

The function $\Ezrho(\varrho)$ also plays an important role in the
study of upper bounds to the reliability function. In fact, the sphere
packing error exponent $\Esp(R)$ is given by \cite{Gallager68}
\begin{equation}
  \label{eq:jazz70}
  \Esp(R) = \max_{\varrho \geq 0} \{ \Ezrho(\varrho) - \varrho R\}.
\end{equation}

Combining \eqref{eq:jazz30} with \eqref{eq:amos104} and
\eqref{eq:amos25} we obtain the two equivalent expressions for
$\Ezrho(\varrho)$
\begin{equation}
  \label{eq:primal_amos}
  \Ezrho(\varrho) =   \max_{\law{Q}} \left\{ 
    - \log \sum_{y \in \Y} 
    \left(\sum_{x\in\X}
      \law{Q}(x) \law{W}(y|x)^\frac{1}{1+\varrho}\right)^{1+\varrho}
                 \right\}
\end{equation}
\begin{equation}
  \label{eq:dual_amos}
  \Ezrho(\varrho) = \max_{\law{Q}} \min_{\law{R}} \left\{ 
    -(1+\varrho) \sum_{x \in \X} \law{Q}(x) \log \left( \sum_{y \in \Y}
      \law{W}(y|x)^{\frac{1}{1+\varrho}}
      \law{R}(y)^{\frac{\varrho}{1+\varrho}} \right) 
                                  \right\}.
\end{equation}
We refer to the former expression as the ``primal'' expression and to
the latter as the ``dual'' expression. 
The primal expression is useful for the derivation of lower bounds on
$\Ezrho(\varrho)$. Indeed, any distribution $\law{Q}$ on the input
alphabet $\set{X}$ induces the lower bound
\begin{equation}
  \Ezrho(\varrho) \geq 
    - \log \sum_{y \in \Y} 
    \left(\sum_{x\in\X}
      \law{Q}(x) \law{W}(y|x)^\frac{1}{1+\varrho}\right)^{1+\varrho}.
\end{equation}
On the other hand, the dual expression is useful for the derivation of
upper bounds. Any distribution $\law{R}$ on the output alphabet
$\set{Y}$ yields the upper bound
\begin{align}
\Ezrho(\varrho) & \leq \max_{\law{Q}} \left\{ 
    -(1+\varrho) \sum_{x \in \X} \law{Q}(x) \log \left( \sum_{y \in \Y}
      \law{W}(y|x)^{\frac{1}{1+\varrho}}
      \law{R}(y)^{\frac{\varrho}{1+\varrho}} \right) 
                                  \right\} \\
                & = \max_{x \in \set{X}} \left\{ -(1+\varrho)  
\log \left( \sum_{y \in \Y}
      \law{W}(y|x)^{\frac{1}{1+\varrho}}
      \law{R}(y)^{\frac{\varrho}{1+\varrho}} \right) \right\}.
\end{align}

\subsection{Constrained Inputs}

Before we can use the above bounds for fading channels we need to
extend the discussion to cost constrained channels and to channels over
infinite input and output alphabets where the method of types cannot be
directly used. For now we continue our assumption of finite alphabets
and address the cost constraint.

Suppose we limit ourselves to blockcode transmissions where we
only allow codewords $(x_{1}, \ldots, x_{n})$ that satisfy 
 \begin{equation}
   \label{eq:block_const}
   \sum_{\ell=1}^{n}g(x_\ell)\leq n \Upsilon
 \end{equation}
 where $g:\set{X} \to \Reals^+$ is a cost function on the input
 alphabet $\set{X}$, $\Upsilon$ is some pre-specified non-negative
 number, and $n$, as before, is the blocklength. The reliability
 function $E(R)$ is defined as in \eqref{eq:amos1} with the
 modification that $\text{P}_{\text{e}}(n, R)$ should be now
 understood as the lowest average probability of error that can be
 achieved using a rate-$R$ blocklength-$n$ codebook all of whose
 codewords satisfy the cost constraint.

To obtain lower bounds on $E(R)$ Gallager \cite{Gallager68},
\cite{Gallager65} modified his random coding
argument in two ways. He introduced a new ensemble of codebooks and
introduced an improved technique to analyze the average probability of
error over this ensemble. For any probability law $\law{Q}$ on the
input alphabet satisfying
\begin{equation}
\label{eq:sat}
  \E[\law{Q}]{g(X)} \leq \Upsilon
\end{equation}
where
\begin{equation}
  \label{eq:def_expectation}
  \E[\law{Q}]{g(X)} \triangleq \sum_{x \in \set{X}} \law{Q}(x) g(x)
\end{equation}
define
\begin{equation}
  \label{eq:def_GalErqM}
  \GalErqM(\varrho, \law{Q}) \triangleq \begin{cases}
    \GalEzrhoq(\varrho, \law{Q}) & \text{if $\E[\law{Q}]{g(X)} <
    \Upsilon$} \\
  {\displaystyle
    \max_{r \geq 0} E_{0}(\varrho, \law{Q}, r)
    } & \text{if
    $\E[\law{Q}]{g(X)} = \Upsilon$} 
  \end{cases}
\end{equation}
where
\begin{equation}
  \label{eq:modGall}
  E_0(\varrho,\law{Q},r) \triangleq - \log \sum_{y \in \Y} 
  \left(\sum_{x\in\X} \law{Q}(x) e^{r(g(x) - \Upsilon)} \law{W}(y|x)^\frac{1}{1+\varrho}
 \right)^{1+\varrho}.
\end{equation}
Note that 
\begin{equation}
  \label{eq:shir1}
  \Bigl. E_0(\varrho,\law{Q},r) \Bigr|_{r=0} = \GalEzrhoq(\varrho,\law{Q}) 
\end{equation}
and hence
\begin{equation}
  \label{eq:shir2}
\max_{r \geq 0} E_{0}(\varrho, \law{Q}, r) \geq   
\GalErqM(\varrho, \law{Q})  \geq \GalEzrhoq(\varrho, \law{Q}).
\end{equation}
Thus, Gallager's ``modification'' can only tighten the bound.

Gallager then showed that for any $0 \leq \varrho \leq 1$ the exponent
\[ \GalErqM(\varrho, \law{Q}) - \varrho R\]
is achievable using block codes that satisfy the
constraint. 

(To prove this result when $\E[\law{Q}]{g(X)} < \Upsilon$ he
considered an ensemble of codebooks where the codewords are chosen
independently of each other, each according to the a-posteriori law of
a sequence $X_{1}, \ldots, X_{n}$ drawn IID according to $\law{Q}$
conditional on $\sum_{k=1}^{n} g(X_{k}) \leq n \Upsilon$. To
prove the result when $\E[\law{Q}]{g(X)} = \Upsilon$ he considered an
ensemble similarly constructed but with the distribution being
conditional on $n \Upsilon - \delta \leq \sum_{k=1}^{n} g(X_{k})
\leq n \Upsilon$.)

Consequently the error exponent
\begin{equation}
  \label{eq:shir5a}
   \GalErM(R, \Upsilon) \triangleq 
  \max_{0 \leq \varrho \leq 1} \left\{ \GalErqM(\varrho, \Upsilon) - 
    \varrho R \right\}
\end{equation}
where
\begin{equation}
  \label{eq:shir5b}
  \GalErqM(\varrho, \Upsilon) \triangleq \max_{\law{Q}:
  \E[\law{Q}]{g(X)} \leq \Upsilon} \GalErqM(\varrho, \law{Q})
\end{equation}
is achievable.

It is instructive to distinguish between two types of constraints. We say
that the cost constraint is \emph{inactive} if there exists some input
distribution $\law{Q}^{*}$ satisfying the constraint that achieves the
global unconstrained maximum of $\GalEzrhoq(\varrho, \law{Q})$. That is, 
\begin{equation}
  \label{eq:inactive1}
  \exists \law{Q}^{*}: \; 
  \E[\law{Q}^{*}]{g(X)} \leq \Upsilon  \quad \text{and} \quad
  \GalEzrhoq(\varrho, \law{Q}^{*}) =  \max_{\law{Q}} \GalEzrhoq(\varrho, \law{Q})
\end{equation}
or equivalently
\begin{equation}
  \label{eq:inactive2}
  \max_{\law{Q}: \E[\law{Q}]{g(X)} \leq \Upsilon} \GalEzrhoq(\varrho,
  \law{Q}) =   \max_{\law{Q}} \GalEzrhoq(\varrho, \law{Q}).
\end{equation}
Otherwise, we say that the cost constraint is \emph{active}. With these
definitions it can be shown that \eqref{eq:shir5b} simplifies to 
\begin{equation}
  \label{eq:max_simple}
  \GalErqM(\varrho, \Upsilon) = \begin{cases}
    {\displaystyle \max_{\law{Q}: \E[\law{Q}]{g(X)} = \Upsilon} \; 
      \max_{r \geq 0} E_0(\varrho,\law{Q},r)} & \text{cost active} \\
  {\displaystyle \max_{\law{Q}} \GalEzrhoq(\varrho, \law{Q})} 
    & \text{cost inactive}
    \end{cases}.
\end{equation}
(The case where the cost constraint is active follows from Gallager's observation
that when the cost constraint is active, the maximum of
$E_{0}(\varrho, \law{Q}, r)$ over all $r \geq 0$ and over all laws
$\law{Q}$ satisfying \eqref{eq:sat} is achieved by an input
distribution $\law{Q}_{*}$ satisfying the constraint with equality.
The case where the cost constraint is inactive follows by noting that by
starting from \eqref{eq:shir2} we have for inactive cost constraints
\begin{align*}
  \max_{\law{Q}:
  \E[\law{Q}]{g(X)} \leq \Upsilon} \GalErqM(\varrho, \law{Q}) & \geq 
\max_{\law{Q}:
  \E[\law{Q}]{g(X)} \leq \Upsilon} \GalEzrhoq(\varrho, \law{Q}) \\
& = \max_{\law{Q}} \GalEzrhoq(\varrho, \law{Q}) \\
& = \max_{\law{Q}} \CKEzrhoq(\varrho, \law{Q}) \\
& \geq \max_{\law{Q}: \E[\law{Q}]{g(X)} \leq \Upsilon}
  \GalErqM(\varrho, \law{Q}) 
\end{align*}
so that all inequalities must hold with equalities. Here the first
inequality follows from \eqref{eq:shir2}; the subsequent equality
because the cost constraint is assumed inactive \eqref{eq:inactive2}; the
subsequent equality from \eqref{eq:jazz30}; and the final inequality
from \eqref{eq:shir3} ahead.)

An achievable error exponent can also be demonstrated using constant
composition codes. This yields that the error exponent
\begin{equation}
  \label{eq:shir6a}
  \CKEr(R, \Upsilon) \triangleq 
  \max_{0 \leq \varrho \leq 1} \left\{ \CKEzrhoq(\varrho, \Upsilon) -
  \varrho R \right\}
\end{equation}
is achievable where
\begin{equation}
  \label{eq:shir6b}
\CKEzrhoq(\varrho, \Upsilon) \triangleq \max_{\law{Q}:
  \E[\law{Q}]{g(X)} \leq \Upsilon} \CKEzrhoq(\varrho, \law{Q}).
\end{equation}

The relation \eqref{eq:shir2} not withstanding, it can be shown that
for any law $\law{Q}$ satisfying \eqref{eq:sat} and any $\varrho \geq 0$
\begin{equation}
  \label{eq:shir3}
  \CKEzrhoq(\varrho, \law{Q}) \geq \GalErqM(\varrho, \law{Q})
\end{equation}
with the inequality being, in general, strict.\footnote{In the case
  $\E[\law{Q}]{g(X)} < \Upsilon$ this follows directly from
  \eqref{eq:jazz20}. For a proof in the case $\E[\law{Q}]{g(X)} =
  \Upsilon$ see Proposition~\ref{prop:upper} ahead, which proves that the
  RHS of \eqref{eq:amos104} is greater or equal $\GalErqM(\varrho,
  \law{Q})$.}  Consequently, by \eqref{eq:shir6b} and
\eqref{eq:shir5b} we have $\CKEzrhoq(\varrho, \Upsilon) \geq \GalErqM(\varrho, \Upsilon)$. However, as shown in
Appendix~\ref{app:equal_rc} this holds with equality
\begin{equation}
  \label{eq:same_const}
  \CKEzrhoq(\varrho, \Upsilon) = \GalErqM(\varrho,
\Upsilon).
\end{equation}
Thus, denoting the two identical functions $\GalErqM(\varrho,
\Upsilon)$ and $\CKEzrhoq(\varrho, \Upsilon)$ by $E_{0}(\varrho,
\Upsilon)$ and the two identical functions $\CKEr(R, \Upsilon)$ and
$\GalErM(R, \Upsilon)$ by $E_{\textnormal{r}}(R, \Upsilon)$ we have
\begin{equation}
  \label{eq:const_rand_exp}
  E_{\textnormal{r}}(R, \Upsilon) = \max_{0 \leq \varrho \leq 1}
  \left\{ E_{0}(\varrho, \Upsilon) - \varrho R \right\} 
\end{equation}
where $E_{0}(\varrho, \Upsilon)$ can be expressed either by (\ref{eq:max_simple}) as
\begin{equation}
  \label{eq:kineret1}
 E_{0}(\varrho, \Upsilon) = \begin{cases}
  {\displaystyle \max_{\law{Q}: \E[\law{Q}]{g(X)} = \Upsilon} \; \max_{r \geq 0}
    E_0(\varrho,\law{Q},r)} & \text{cost active} \\
    {\displaystyle \max_{\law{Q}} \GalEzrhoq(\varrho, \law{Q})} 
    & \text{cost inactive}
    \end{cases}
\end{equation}
or, using \eqref{eq:amos104}, as
\begin{multline}
  \label{eq:kineret2}
  E_{0}(\varrho, \Upsilon) = \\ \max_{\law{Q}: \E[\law{Q}]{g(X)} \leq
    \Upsilon} 
  \min_{\law{R}} \left\{ 
    -(1+\varrho) \sum_{x \in \X} \law{Q}(x) \log \left( \sum_{y \in \Y}
      \law{W}(y|x)^{\frac{1}{1+\varrho}}
      \law{R}(y)^{\frac{\varrho}{1+\varrho}} \right) 
                                  \right\}.
\end{multline}
The former, to which we refer as the ``primal'' expression, is useful
for the derivation of lower bounds on $E_{0}(\varrho, \Upsilon)$
whereas the latter, the ``dual'', is useful for upper bounds.

\section{Continuous Alphabets \label{sec:continuous}}

We next extend the discussion to channels over infinite input and
output alphabets. Consider a channel $W(\cdot|\cdot)$ whose inputs and
outputs take value in the separable metric spaces $\X$ and $\Y$
respectively. Thus for any input $x \in \X$ and any Borel set $\set{B}
\subset \Y$ the probability that in response to the input $x$ the
channel will produce an output $Y$ that lies in the set $\set{B}$ is
$W(\set{B}|x)$. We assume that the mapping $x \mapsto W(\set{B}|x)$
from $\set{X}$ to the interval $[0,1]$ is Borel measurable. Finally
assume the existence of an underlying positive measure $\law{\mu}$ on $\Y$
with respect to which all the probability measures $\{W(\cdot|x), x
\in \set{X}\}$ are absolutely continuous. Denote the Radon-Nykodim
derivative of $W(\cdot|x)$ with respect to $\mu$ by
\[ w(\cdot|x) = \frac{\d W(\cdot|x)}{\d \mu},\quad x \in \X.\]
Thus, $w(y|x)$ is the density at $y$ of the channel
output corresponding to the input $x \in \X$. For any input $x \in \set{X}$
and any Borel set $\set{B} \subset \Y$
\begin{equation}
  W(\set{B}|x) = \int_{\set{B}} w(y|x) \d{\mu}(y).
\end{equation}

As to the cost, we shall assume that the function $g: \X \rightarrow
\Reals^{+}$ is measurable and consider block codes that satisfy
\eqref{eq:block_const}. We extend the definition
\eqref{eq:def_expectation} to infinite alphabets as
\begin{equation}
  \label{eq:def_avg_cost}
  \E[\law{Q}]{g(X)} \triangleq \int_{\X} g(x) \d \law{Q}(x).
\end{equation}

Definition~\eqref{eq:modGall} is extended for any probability law
$\law{Q}$ on $\X$ as
\begin{equation}
\label{eq:modGallCont}
   E_0(\varrho, \law{Q}, r) \triangleq - \log \int_{y \in \Y} 
   \left(\int_{x \in \X}e^{r(g(x) -  \Upsilon)}w(y|x)^\frac{1}{1+\varrho} \d
    \law{Q}(x)\right)^{1+\varrho}\d \mu(y).
\end{equation}

For any input distribution $\law{Q}$ satisfying the constraint
$\E[\law{Q}]{g(X)} \leq \Upsilon$ we extend \eqref{eq:def_GalErqM} as
follows:
\begin{equation}
  \label{eq:mod_infinite}
    \GalErqM(\varrho, \law{Q}) \triangleq \begin{cases}
  {\displaystyle
    \sup_{r \geq 0} E_{0}(\varrho, \law{Q}, r)
    } & \text{if
    $\E[\law{Q}]{g(X)} = \Upsilon$ and $\E[\law{Q}]{g^{3}(X)} <
  \infty$} \\
{\displaystyle
  \Bigl. E_{0}(\varrho, \law{Q}, r) \Bigr|_{r=0}} &  \text{otherwise}
\end{cases}.
\end{equation}
(Note that following Gallager \cite{Gallager68}, \cite{Gallager65} we
allow for the optimization over $r$ only when under the law $\law{Q}$
the random variable $g(X)$ has a finite third moment.)

With this definition we can now define 
\begin{equation}
  \label{eq:def_Ezcont}
  E_{0}(\varrho, \Upsilon) \triangleq \sup_{\law{Q}: \E[\law{Q}]{g(X)} \leq
  \Upsilon} \GalErqM(\varrho, \law{Q})
\end{equation}
and the  cut-off rate as
\begin{equation}
  \label{eq:Rzero_cont}
  R_{0}(\Upsilon) \triangleq \Bigl. E_{0}(\varrho, \Upsilon) \Bigr|_{\varrho = 1}. 
\end{equation}
The random coding error exponent 
\begin{equation*}
 \sup_{0 \leq \varrho \leq 1} \left\{
E_{0}(\varrho, \Upsilon) - \varrho R \right\}
\end{equation*}
is achievable with block codes satisfying the constraint
\eqref{eq:block_const} \cite{Gallager68}, \cite{Gallager65}.

The following proposition proves \eqref{eq:shir3} in the more general
case where the alphabets may be continuous. It is particularly useful
for the derivation of upper bounds on $\GalErqM(\varrho, \Upsilon)$.
\begin{proposition}
\label{prop:upper}
Consider as above a discrete-time memoryless infinite  alphabet
channel $w(y|x)$, an output measure $\mu$, a measurable cost function
$g: \X \rightarrow \Reals^{+}$, and some arbitrary allowed cost
$\Upsilon$. Let $f_{R}$ be an arbitrary density with respect to $\mu$
on the output alphabet $\Y$. Then for any distribution $\law{Q}$ on
$\X$ satisfying the cost constraint $\E[\law{Q}]{g(X)} \leq \Upsilon$
\begin{multline}
  \label{eq:upper_cont}
  \GalErqM(\varrho, \law{Q}) \leq \\ -(1+\varrho) \int_{x \in \X} \log
 \left( \int_{y \in \Y} w(y|x)^{\frac{1}{1+\varrho}} 
   f_{R}(y)^{\frac{\varrho}{1+\varrho}}
      \d \mu(y) \right) \d \law{Q}(x).
\end{multline}
\end{proposition}
\begin{proof}
  Distinguish between the case where $\E[\law{Q}]{g(X)} < \Upsilon$
  and the case where $\E[\law{Q}]{g(X)} = \Upsilon$ and
  $\E[\law{Q}]{g^{3}(X)} < \infty$. In the former case, by
  \eqref{eq:mod_infinite}, $\GalErqM(\varrho, \law{Q}) =
  E_{0}(\varrho, \law{Q}, 0)$ and the result follows by an application
  of Jensen's inequality and H\"{o}lder's inequality:
\begin{eqnarray*}
  \lefteqn{-(1+\varrho) \int_{x \in \X} \log \left( \int_{y \in
        \Y} w(y|x)^\frac{1}{1+\varrho} f_{R} (y)^\frac{\varrho}{1+\varrho}
      \d \mu(y) \right)\d \law{Q}(x)} \nonumber\\ 
  & \geq & -(1+\varrho) \log \int_{x \in \X} \int_{y \in \Y} 
  w(y|x)^\frac{1}{1+\varrho} f_{R}(y)^\frac{\varrho}{1+\varrho} 
  \d \mu(y)\d \law{Q}(x) \nonumber\\
  & = & -(1+\varrho) \log \int_{y \in \Y}  \left( \int_{x \in \X}
        w(y|x)^\frac{1}{1+\varrho} \d \law{Q}(x) \right) \cdot 
      \left( f_{R}(y)^\frac{\varrho}{1+\varrho} \right) \d \mu(y) \\
  & \geq &  - \log \int_{y \in \Y} \left(\int_{x \in \X}
        w(y|x)^\frac{1}{1+\varrho}  \d \law{Q}(x)\right)^{1+\varrho}\d
        \mu(y) \\
  & = & E_{0}(\varrho, \law{Q}, 0).
\end{eqnarray*}

As for the case where $\E[\law{Q}]{g(X)} = \Upsilon$ (and
$\E[\law{Q}]{g^{3}(X)} < \infty$) we have for any $r \geq 0$
\begin{eqnarray}
  \lefteqn{-(1+\varrho)\int_{x \in \X} \log \left( \int_{y \in
        \Y} w(y|x)^\frac{1}{1+\varrho} f_{R}(y)^\frac{\varrho}{1+\varrho}
      \d \mu(y) \right) \d \law{Q}(x)} \nonumber\\ 
  & = & r(1+\varrho) \left( \E[\law{Q}]{g(X)} - \Upsilon \right)  \nonumber \\
  && -\: (1+\varrho)\int_{x \in \X}\log 
  \left( \int_{y \in \Y} e^{r(g(x)-\Upsilon)}
        w(y|x)^\frac{1}{1+\varrho}
        f_{R}(y)^\frac{\varrho}{1+\varrho} \d \mu(y) \right)\d
        \law{Q}(x) 
        \nonumber \\ 
  & = & -(1+\varrho) \int_{x \in \X} \log \left(\int_{y \in \Y} 
    e^{r(g(x)-\Upsilon)} w(y|x)^\frac{1}{1+\varrho} 
    f_{R}(y)^\frac{\varrho}{1+\varrho} \d \mu(y) \right)\d \law{Q}(x) 
  \nonumber \\ 
  & \geq & -(1+\varrho) \log \int_{x \in \X} \int_{y \in \Y} 
  e^{r(g(x)-\Upsilon)} w(y|x)^\frac{1}{1+\varrho} 
  f_{R}(y)^\frac{\varrho}{1+\varrho} \d \mu(y)\d \law{Q}(x) 
  \nonumber\\ 
  & \geq &  - \log \int_{y \in \Y} \left( \int_{x \in \X} 
    e^{r(g(x) -   \Upsilon)} w(y|x)^\frac{1}{1+\varrho} \d
    \law{Q}(x) \right)^{1+\varrho} \d \mu(y) \nonumber \\
  & = & E_{0}(\varrho, \law{Q}, r).
\end{eqnarray}
where the second equality follows because in the case we are
considering now $\E[\law{Q}]{g(X)} = \Upsilon$; the first inequality
by Jensen's inequality, and the subsequent by H\"older's
inequality. The result for this case now follows because $r \geq 0$ in
the above is arbitrary.
\end{proof}

To conclude, to derive lower bounds on $E_{0}(\varrho, \Upsilon)$ we
can choose any input distribution $\law{Q}$ satisfying the constraint
$\E[\law{Q}]{g(X)} \leq \Upsilon$ to obtain the lower bound:
\begin{equation}
  \label{eq:cont_lb}
  E_{0}(\varrho, \Upsilon) \geq \GalErqM(\varrho, \law{Q}) 
\end{equation}
where $\GalErqM(\varrho, \law{Q})$ is defined in \eqref{eq:mod_infinite}.

To derive upper bounds on $E_{0}(\varrho, \Upsilon)$ we can use the
above proposition by choosing some arbitrary output density $f_{R}(y)$
to obtain
\begin{multline}
  \label{eq:cont_ub}
  E_{0}(\varrho, \Upsilon) \leq \\
  \sup_{\law{Q}: \E[\law{Q}]{g(X)} \leq
  \Upsilon}\left\{ -(1+\varrho) \int_{x \in \X} \log
 \left( \int_{y \in \Y} w(y|x)^{\frac{1}{1+\varrho}} 
   f_{R}(y)^{\frac{\varrho}{1+\varrho}}
      \d \mu(y) \right) \d \law{Q}(x)\right\}.
\end{multline}

\section{Ricean Fading Channels \label{sec:Ricean}}

The discrete-time memoryless Ricean fading channel with partial
receiver side information is a channel whose input $x$ takes value in
the complex field $\Complex$ and whose corresponding output
constitutes of a pair of complex random variables $Y$ and $S$. We
shall refer to $Y$ as ``the received signal'' and to $S$ as the ``side
information (at the receiver)''. The joint distribution of $Y, S$
corresponding to the input $x \in \Complex$ is best described using
the fading complex random variable $H$ and the additive noise complex
random variable $Z$. 

The joint distribution of $H$, $S$, and $Z$ does not depend on the
input $x$. The additive noise $Z$ is independent of the pair $(H,S)$ and
has a circularly symmetric complex Gaussian distribution of positive
variance $\sigma^{2}$.  The fading $H$ is of mean $d \in \Complex$ ---
the ``specular component'' --- and it is assumed that $H - d$ is a
unit-variance circularly symmetric complex Gaussian random variable.\footnote{We shall
  sometimes refer to such Ricean fading as ``normalized Ricean
  fading'' to make it explicit that the fading is of unit variance.
  ``Un-normalized'' Ricean fading need not have unit-variance. Those
  can be normalized by scaling the fading and absorbing the scaling
  into the input power. Note also that there is no loss in generality
  in assuming that $d$ is real and non-negative. The more general
  complex case can be treated by rotating the output.} The pair $S$
and $H-d$ are jointly circularly symmetric Gaussian random variables. We denote the
conditional variance of $H$ given $S$ by $\eps^{2}$.

The received signal $Y$ corresponding to the input $x \in \Complex$ is
given by
\begin{equation}
  \label{eq:channel_law}
  Y = H x + Z.
\end{equation}

The case where $\eps^{2} = 1$ corresponds to the case where
$H$ and $S$ are independent, in which case the receiver can discard $S$
without loss in information rates. This case corresponds to
``non-coherent'' fading. In the case $\eps^{2} = 0$ the receiver can
precisely determine the realization of $H$ from $S$. This corresponds
to ``coherent detection''. Finally, the case $0 < \eps < 1$ corresponds
to ``partially coherent'' communication. In this case $S$ carries some
information about $H$, but it does not fully determine $H$. In this
paper we shall only consider the case where $\eps^{2} > 0$.  The case
$\eps^{2} = 0$ is much easier to analyze and has already received
considerable attention in the literature. See for example,
\cite{Ericson70}, \cite{Ahmed_McLane99},
\cite{Biglieri_Proakis_Shamai98} and the references in the latter.

The special case of Ricean fading with zero specular component $d$ is
called ``Rayleigh fading''. The non-coherent ($\eps^{2} = 1$) capacity
of this channel was studied in \cite{Faycal_Trott_Shamai},
\cite{Taricco_Elia97} and \cite{Lapidoth_Moser03}. The coherent case
($\eps^{2} = 0$) was studied in \cite{Ericson70}.  The capacity of the
non-coherent Ricean channel ($\eps^{2} = 1$ and $d \neq 0$) was
studied in \cite{Gursoy_Poor_Verdu1}-\cite{Gursoy_Poor_Verdu2} and
\cite{Lapidoth_Moser03}.

Unless some restrictions are imposed on the input $x$, the capacity
and cut-off rate of this channel are infinite. Two kinds of
restrictions are typically considered. The first corresponds to an
average power constraint. Here only blockcodes where each
codeword satisfies \eqref{eq:block_const} with
\begin{equation}
  \label{eq:def_g}
  g(x) = |x|^{2}
\end{equation}
are allowed. In this context rather than denoting the allowed cost by
$\Upsilon$ we shall use the more common symbol $\Es$, which stands
here for the average energy per symbol. That is, we only allow
blocklength-$n$ codes in which every codeword $x_{1}, \ldots, x_{n}$
satisfies
\begin{equation}
  \label{eq:avg_const}
  \frac{1}{n} \sum_{\ell=1}^{n} |x_{\ell}|^{2} \leq \Es.
\end{equation}

The second type of constraint is a peak power constraint. Here we
only allow channel inputs that satisfy
\begin{equation}
  \label{eq:peak_const}
  |x|^{2} \leq \Es
\end{equation}
where $\Es$ now stands for the allowed peak power. Such a constraint
is best treated by considering the channel as being free of
constraints but with the input alphabet now being $\{z \in \Complex:
|z|^{2} \leq \Es\}$.

For both the average and peak power constraints we define the
signal-to-noise ratio (SNR) as
\begin{equation}
  \label{eq:def_SNR}
  \textnormal{SNR} \triangleq \frac{\Es}{\sigma^{2}}.
\end{equation}

Any codebook satisfying the peak power constraint
\eqref{eq:peak_const} also satisfies the average power constraint
hence the capacity and reliability function under the peak constraint
cannot exceed those under the average constraint.

Irrespective of whether an average power or a peak power constraint is
imposed, at high SNR the capacity $C(\textnormal{SNR}|S)$ of this
channel is given asymptotically as
\begin{equation}
  \label{eq:capacity_Rice}
  C(\textnormal{SNR}|S) = \log \log \textnormal{SNR} + \log \dabs^{2}
  - \textnormal{Ei}\bigl(-\dabs^{2}\bigr) -1 + \log \frac{1}{\eps^{2}} + o(1)
\end{equation}
where the correction term $o(1)$ depends on the SNR and tends to zero
as the SNR tends to infinity.  Here $\textnormal{Ei}(\cdot)$ denotes the
Exponential Integral function
\begin{equation}
  \label{eq:def_Ei}
  \textnormal{Ei}(-\xi) = -\int_{\xi}^{\infty} \frac{e^{-t}}{t} \d{t}, 
  \qquad \xi > 0
\end{equation}
and we define the value of the function $\log(\xi) - \textnormal{Ei}(-\xi)$
at $\xi = 0$ as $-\gamma$, where $\gamma \approx 0.577$ denotes
Euler's constant. (With this definition the function $\log(\xi) -
\textnormal{Ei}(-\xi)$ is continuous from the right at $\xi = 0$.)

Here we shall study the cutoff rate in two cases.  First, in the
absence of side information ($\eps^{2} = 1$) we will show that
irrespective of whether a peak or average power constraint is imposed
\begin{multline}
  \label{eq:cutoff_noSI}
    R_{0}(\textnormal{SNR}) = \log \log \textnormal{SNR} + 
        \frac{\dabs^2}{2} - \log(2 \pi) - 2 \log \I0
        \left(\frac{\dabs^2}{4}\right) + o(1).
\end{multline}
Here $\I0(\cdot)$ denotes the zero-th order modified Bessel function
of the first kind, which is given by
\begin{equation}
  \label{eq:def_I0}
  \I0(\xi) = \frac{1}{2 \pi} \int_{-\pi}^{\pi} e^{\xi \cos \theta} \d
  \theta, \qquad \xi \in \Reals 
\end{equation}
and the $o(1)$ term is a correction term that depends on the SNR and
that approaches zero as the SNR tends to infinity.

\begin{figure}
  \centering \psfrag{sot}[cc][cc]{Second Order Terms (nats)}
  \psfrag{d}[cc][bc]{The specular component $\dabs$}
  \psfrag{data12345}[cc][cc]{\tiny of $R_0(\textnormal{SNR})$}
  \psfrag{data23456}[cc][cc]{\tiny of $C(\textnormal{SNR})$}
  \psfrag{dataaaaaaa123456789}[cc][cc]{\tiny $\lim
    \{(C-R_0)(\textnormal{SNR})\}$} \epsfig{file=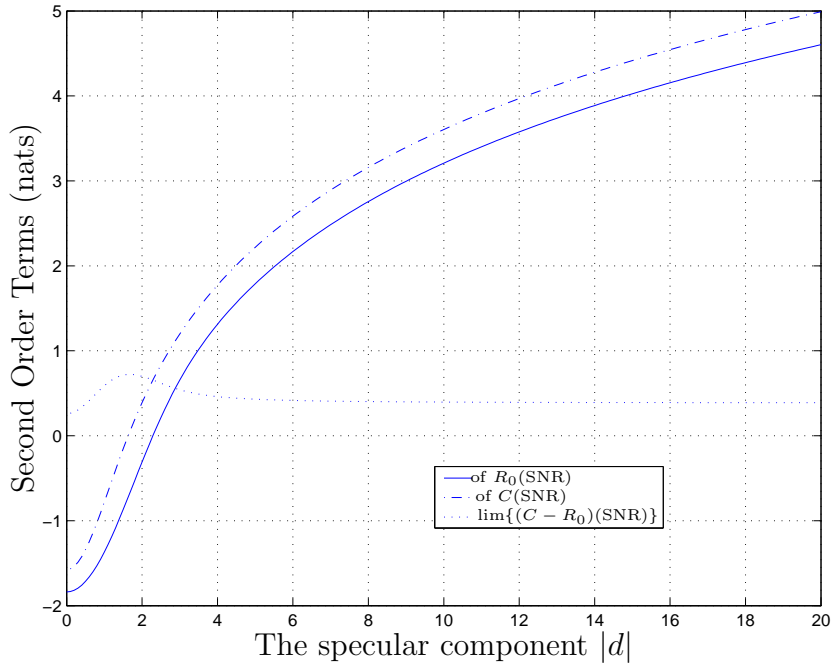,
    width=0.8\textwidth}
  \caption{The second order terms of 
    $C(\textnormal{SNR})$ and $R_0(\textnormal{SNR})$ and their
    difference as functions of the specular component $\dabs$ for
    $\eps = 1$, i.e., in the absence of side information. Upper curve
    depicts $\displaystyle{\lim_{\textnormal{SNR} \rightarrow \infty}
      \{ C(\textnormal{SNR}) - \log \log \textnormal{SNR}\}}$,
    followed by the analogous term for the cutoff rate and their
    difference.}
  \label{fig:plot1}
\end{figure}

Figure~\ref{fig:plot1} depicts the second order term (the constant
term) in the high SNR expansion of channel capacity
\eqref{eq:capacity_Rice} and of the  cutoff rate
\eqref{eq:cutoff_noSI} as a function of the specular component $d$ in
the absence of side information. For a zero specular component the
difference between the two second order terms is $\log (2 \pi) - 1
-\gamma \approx 0.26$ nats; for very large specular components ($\dabs
\rightarrow \infty$) this difference approaches $\log(4/e) \approx
0.39$ nats.\footnote{All logarithms in this paper are natural
  logarithms.} 

For the case where the side information is present but is not perfect
($0 < \eps^{2} < 1$) we only treat the case of zero specular component
($d = 0$, i.e., Rayleigh fading). We obtain the expansion
\begin{multline}
  \label{eq:cutoff_with_si}
    R_{0}(\textnormal{SNR}|S) = \log \log \textnormal{SNR} +
        \log\frac{1}{\epsilon^2} - 
        \log \K \left( \sqrt{1-\epsilon^4}\right) -\log4 + o(1)
    \\
    0 < \eps^{2} < 1, \quad d=0
\end{multline}
where $\K(\cdot)$ is the complete elliptic integral of the first kind:
\begin{equation}
  \label{eq:def_K}
  \K(\xi) = \int_{0}^{1} \frac{1}{\sqrt{1-t^{2}} \sqrt{1 - \xi^{2}
  t^{2}}} \d t, \qquad \xi^{2} < 1.
\end{equation}

For the case of Rayleigh fading with perfect side information
($\eps^{2} = 0$) see \cite{Ahmed_McLane99}. For the case of ``almost
perfect side information'' ($0 < \eps^{2} \ll 1$) we note the
expansion
\begin{multline}
         \log\frac{1}{\epsilon^2} - 
        \log \K \left( \sqrt{1-\epsilon^4}\right) -\log 4
        = \log\frac{1}{\epsilon^2} -\log\log\frac{4}{\eps^{2}} - \log4
        + o(\epsilon^4)
        \\
         \quad 0 < \eps^{2} \ll 1.
\end{multline}
which 
follows from the approximation \cite{Carlson_Gustafson85}
\begin{equation}
  \label{eq:carlson1}
  \K(k) = \frac{1}{1-\theta} \log \frac{4}{\sqrt{1-k^{2}}}, \quad 0
  \leq k < 1
\end{equation}
for some 
\begin{equation}
  \label{eq:carlson2}
  0 < \theta < \frac{1 - k^{2}}{4}.
\end{equation}

\begin{figure}
  \label{fig:plot}
  \centering \psfrag{sot}[cc][cc]{Second Order Terms (nats)}
  \psfrag{eps}[cc][bc]{The Estimation Error $\epsilon^2$}
  \psfrag{data234567}[cc][cc]{\tiny of $R_0(\textnormal{SNR})$}
  \psfrag{data123456}[cc][cc]{\tiny of $C(\textnormal{SNR})$}
  \psfrag{data345678901234567}[cc][cc]{\tiny $\lim
    \{(C-R_0)(\textnormal{SNR})\}$} \epsfig{file=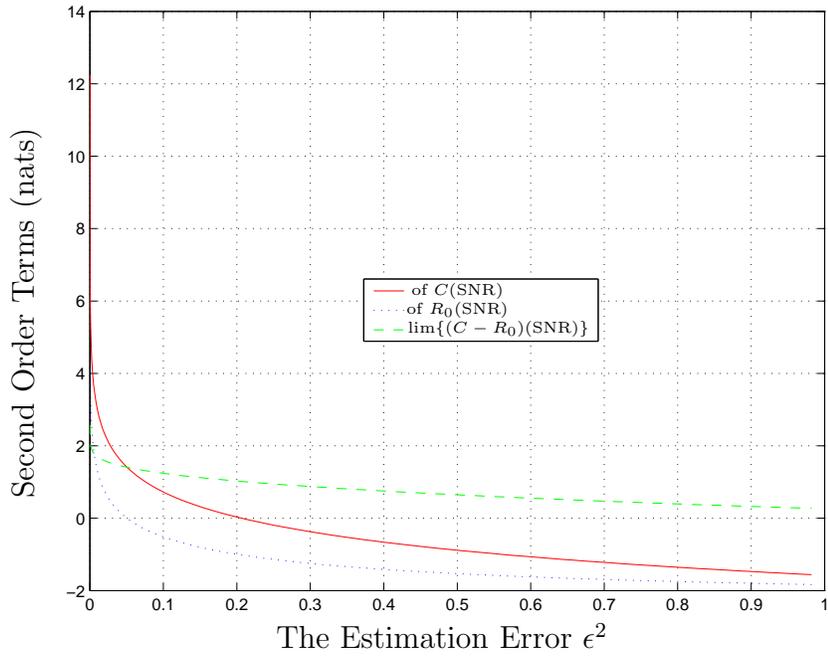,
    width=0.8\textwidth}
  \caption{The second order terms of $C(\textnormal{SNR})$,
    $R_0(\textnormal{SNR})$ and their difference as functions of the
    minimum mean squared error $\eps^{2}$ in estimating the fading
    from the side information.  Rayleigh fading ($d=0$) is assumed.}
  \label{fig:plot2}
\end{figure}

Figure~\ref{fig:plot2} depicts the second order terms of channel
capacity \eqref{eq:capacity_Rice} and the  cutoff rate
\eqref{eq:cutoff_with_si} as a function of the estimation error
$\eps^{2}$ in estimating the
fading from the side information for Rayleigh fading channels ($d=0$).

\section{Derivations for Ricean Channels \label{sec:derivation}}
\subsection{The Cut-Off Rate in Absence of Side Information}
\subsubsection{Upper Bound}
\label{sec:upbound}
To derive an upper bound on the cut-off rate of the Ricean channel in
the absence of side information we use Proposition~\ref{prop:upper}
with the density (w.r.t.\ the Lebesgue measure $\mu$ on $\Complex$)
\begin{equation}
  \label{eq:choice_fR}
  f_{R}(y) =
\frac{(|y|^2 + \delta)^{\alpha-1} e^{-\frac{|y|^2+\delta}
    {\beta}}}{\pi \beta^\alpha \Gamma(\alpha,\delta / \beta)}, \ y \in
\Complex.
\end{equation}
Here the parameters $\delta \geq 0$, $\alpha > 0$, and $\beta > 0$ can
be chosen freely in order to obtain the tightest bound, and $\Gamma
(\alpha, \xi)$ denotes the incomplete Gamma function, 
\begin{equation}
  \label{eq:incomplete_Gamma}
  \Gamma(\alpha, \xi) = \int_{\xi}^{\infty}t^{\alpha -1}e^{-t}\d t, \
  \alpha >0, \xi \geq 0.
\end{equation}
(This family of densities was introduced in \cite{Lapidoth_Moser03}
for the purpose of studying the fading number.)

By Proposition~\ref{prop:upper} applied with $\varrho = 1$ we obtain
for any law $\law{Q}$ under which 
\begin{equation}
  \label{eq:leq_power}
  \E[\law{Q}]{|X|^2} \leq \Es
\end{equation}
the upper bound
\begin{equation}
  \label{eq:lab_star}
    \GalErqM(1, \law{Q}) \leq - 2 \int_{x \in \Complex} \log
    \psi(x) \d \law{Q}(x)
\end{equation}
where
\begin{align}
\psi(x) & \triangleq \int_{y \in \Complex} \sqrt{w(y|x) \cdot
  f_{R}(y)} \d \mu(y) \label{eq:lab_A} \\
& = \frac{2 e^{\frac{-\delta}{2 \beta}}
    e^{-\frac{\dabs^2 |x|^2}{2 (|x|^2 + \sigma^2)}}}
  {\sqrt{\Gamma(\alpha, \frac{\delta}{\beta})}
  \beta^{\frac{\alpha}{2}} \sqrt{|x|^2+\sigma^2}} \ell(x; \alpha,
  \beta, \delta) \label{eq:lab_B}
\end{align}
and from $\cite[3.338]{Grad_Ryz_94}$
\begin{equation}
  \ell(x; \alpha, \beta, \delta) = \int_{0}^{\infty}
    e^{-\frac{\rho^{2} (\beta+|x|^2 + \sigma^2)}
      {2 \beta (|x|^2+\sigma^2)}} \rho (\rho^2+\delta)^\frac{\alpha-1}{2}
    \I0\left( \frac{\dabs \cdot |x| \cdot \rho}{|x|^2+\sigma^2} \right) \d \rho.
\end{equation}


For our high SNR analysis it will suffice to consider (for sufficiently
large powers $\Es$) the possibly sub-optimal choice of the parameters
\begin{equation}
  \label{eq:lab_C}
  \beta = \Es \log \Es \qquad \alpha = \frac{\delta}{\log \beta}
\end{equation}
and to consider the limiting behavior of the bound as $\Es \rightarrow
\infty$. After taking this limit with $\delta > 0$ held fixed we shall
consider the additional limit of $\delta \rightarrow 0$. 

The analytic computation of $\ell(x; \alpha, \beta, \delta)$ is
difficult. Note, however, that any lower bound to this quantity will
yield an upper bound on $\GalErqM(1, \law{Q})$. Also, the integral is
computable when both $\alpha$ and $\delta$ are formally set to
zero.\footnote{In fact, it suffices that $\delta$ be set to zero.} We
can thus use a limiting argument to study $\ell(x; \alpha, \beta,
\delta)$ for $\alpha, \delta$ very small. Indeed, in
Appendix~\ref{app:mim} it is shown that
\begin{equation}
  \label{eq:Mim1}
  \ell(x; \alpha, \beta, \delta) \geq  a(\alpha, \beta, \delta, m_{1})
  \cdot  \ell(x; \alpha = 0, \beta, \delta = 0)
\end{equation}
where
\begin{equation}
  \label{eq:Mim2}
  a(\alpha, \beta, \delta, m_{1}) = \delta^{\alpha/2} 
  \sqrt{\frac{m_1}{m_1+1}} \left( 1 -
    \frac{\sqrt{m_1\delta} \cdot \I0
      \left(\frac{\dabs\sqrt{m_1\delta}}{2\sigma}\right)}
    {\sqrt{\frac{ \pi \beta\sigma^2}{2(\beta+\sigma^2)}}}\right)
\end{equation}
$m_1 >0$ being some constant.
As we shall see, the term $a(\alpha, \beta, \delta, m_{1})$ will have
a negligible asymptotic contribution to our bound. 

The term $\ell(x; \alpha = 0, \beta, \delta = 0)$ can be computed
analytically \cite[6.618]{Grad_Ryz_94}:
\begin{multline}
\label{eq:lab_G3}
\ell(x; \alpha = 0, \beta, \delta = 0) 
  = \sqrt{\frac{\pi}{2}}\sqrt{\frac{\beta(|x|^2+\sigma^2)}{\beta+|x|^2+\sigma^2}}
  \\ \cdot 
  e^{\frac{\beta \dabs^2
  |x|^2}{4(|x|^2+\sigma^2)(\beta+|x|^2+\sigma^2)}} 
  \I0\left( \frac{\beta\dabs^2|x|^2}{4(|x|^2+\sigma^2)(\beta+|x|^2+\sigma^2)}\right).
\end{multline}

We thus conclude from (\ref{eq:lab_star}) , (\ref{eq:lab_B}),
\eqref{eq:Mim1}, and \eqref{eq:lab_G3} 


\begin{IEEEeqnarray*}{rCl}
\IEEEeqnarraymulticol{3}{l}{
\GalErqM(1, \law{Q}) \leq \frac{\delta}{\beta} - 2 \log a(\alpha,
   \beta, \delta, m_{1}) + \alpha \log \beta} \nonumber \\ \quad
   & + & \log \Gamma \left(\alpha,\frac{\delta}{\beta} \right) -
   \log(2 \pi) \nonumber \\
   & + & \E[\law{Q}]{\log \left( 1 + \frac{|X|^2 +
         \sigma^2}{\beta} \right)} \nonumber \\
   & + & \dabs^{2} \E[\law{Q}]{ \frac{|X|^2}{|X|^2+\sigma^2} \cdot
     \left( 1 - \frac{\beta}{\beta + |X|^{2} + \sigma^{2}} \right)}
   \nonumber \\
   & + & \E[\law{Q}]{ \frac{\dabs^{2}}{2}
     \frac{|X|^2}{|X|^2+\sigma^2}\frac{\beta}{\beta+|X|^2+\sigma^2} -
     2 \log \I0 \left(
       \frac{\dabs^2}{4}\frac{|X|^2}{|X|^2+\sigma^2}\frac{\beta}{\beta+|X|^2+\sigma^2}
     \right)}.
\end{IEEEeqnarray*}

 The expectations in the above cannot be computed without knowledge of
 the law $\law{Q}$. We thus proceed to upper bound the expectations
 using the average power constraint \eqref{eq:leq_power}. The expectation of the logarithm is
 upper bounded using Jensen's inequality and the power constraint
 \eqref{eq:leq_power}; the following expectation is upper
 bounded using the point-wise upper bound $|x|^{2}/(|x|^{2} +
 \sigma^{2}) < 1$, Jensen's inequality, and the power constraint \eqref{eq:leq_power}; and
 the final expectation by noting that the function $\xi \mapsto \xi -
 2 \log \I0( \xi/2)$ is monotonically increasing and by noting that
 \begin{equation*}
   \frac{\dabs^{2}}{2}
   \frac{|X|^2}{|X|^2+\sigma^2}\frac{\beta}{\beta+|X|^2+\sigma^2} < \frac{\dabs^{2}}{2}.
 \end{equation*}

We thus conclude that with the allowed average power $\Es$ the cut-off
rate satisfies:
\begin{align*}
R_{0}(\Es) - \log \log \frac{\Es}{\sigma^2} & \leq \frac{\delta}{\beta} - 2 \log a(\alpha,
   \beta, \delta, m_{1}) + \alpha \log \beta \\
  & \quad + \log \Gamma \left(\alpha,\frac{\delta}{\beta} \right) -
  \log \log \frac{\Es}{\sigma^2} \\ 
  & \quad + \log \left( 1 + \frac{\Es + \sigma^2}{\beta} \right) \\
  & \quad + (1 - \frac{\beta}{\beta + \Es + \sigma^{2}})|d|^2   \\
  & \quad + \frac{\dabs^{2}}{2} - 2 \log \I0 \left( \frac{\dabs^{2}}{4}
  \right) - \log (2 \pi). \\
\end{align*}
Holding $\delta > 0$ (small) and $m_{1} > 0$ (large) fixed, and
letting $\Es \rightarrow \infty$ with $\alpha = \alpha(\Es)$ and
$\beta = \beta(\Es)$ as in \eqref{eq:lab_C} we obtain from the above
and \eqref{eq:Mim2}
\begin{align*}
  \varlimsup_{\Es \rightarrow \infty} \{ R_{0}(\Es) - \log \log \frac{\Es}{\sigma^2} \}
  & \leq   \log\left(\frac{m_1+1}{m_1}\right) 
  - 2 \log \left( 1 - \frac{\sqrt{m_1 \delta} \cdot  
      \I0 \left(\frac{\dabs \sqrt{m_1 \delta}}{2 \sigma} \right)}
        {\sqrt{\frac{\pi \sigma^2}{2}}}\right) \nonumber \\
      & \quad + \log \frac{1 - e^{-\delta}}{\delta} \nonumber \\
      & \quad + \frac{\dabs^{2}}{2} - 2 \log \I0 \left( \frac{\dabs^{2}}{4}
  \right) - \log (2 \pi)
\end{align*}
where in computing the limiting difference between the Incomplete
Gamma function and $\log \log \Es$ we used \cite[Appendix
XI]{Lapidoth_Moser03}. Holding $m_{1}$ fixed and letting $\delta
\rightarrow 0$ we obtain
\begin{align*}
  \varlimsup_{\Es \rightarrow \infty} \{ R_{0}(\Es) - \log \log \frac{\Es}{\sigma^2} \}
  & \leq   \log\left(\frac{m_1+1}{m_1}\right) \nonumber \\
        & \quad + \frac{\dabs^{2}}{2} - 2 \log \I0 \left( \frac{\dabs^{2}}{4}
  \right) - \log (2 \pi).
\end{align*}
Letting now $m_{1}$ tend to infinity we obtain the desired asymptotic
upper bound
\begin{equation}
  \label{eq:desired_ub}
  \varlimsup_{\Es \rightarrow \infty} \{ R_{0}(\Es) - \log \log \frac{\Es}{\sigma^2} \}
  \leq \frac{\dabs^{2}}{2} - 2 \log \I0 \left( \frac{\dabs^{2}}{4}
  \right) - \log (2 \pi).
\end{equation}

\subsubsection{Lower Bound \label{sec:lowbound}}

Any input distribution satisfying the cost constraint (possibly
strictly) induces a lower bound on the cut-off rate
\eqref{eq:Rzero_cont}. Indeed, for any input distribution
$\tilde{\law{Q}}$ satisfying the cost constraint
\begin{align}
  R_{0}(\Upsilon) & \geq \left. \GalErqM(\varrho, \tilde{\law{Q}})
  \right|_{\varrho=1} \\
  & \geq \left. E_0(\varrho, \tilde{\law{Q}}, r) \right|_{\varrho =1, r = 0}
\end{align}
where the first inequality follows by the definition of the cut-off
rate \eqref{eq:Rzero_cont} (and holds with equality if
$\tilde{\law{Q}}$ achieves the cut-off rate) and where the second
inequality follows from \eqref{eq:mod_infinite} (and holds with
equality if $\tilde{\law{Q}}$ satisfies the cost constraint with
strict inequality).

We thus proceed to lower bound $E_{0}(1, \tilde{\law{Q}}, 0)$ for a law
$\tilde{\law{Q}}$ of our choice. Under this law, $X$ is a circularly symmetric
random variable with
\begin{equation}
  \label{eq:simon110}
  \log |X|^{2} \sim \text{Uniform} 
\left( \log \log \Es, \log \Es \right).
\end{equation}
The motivation for using this law is that it is known to achieve the
asymptotic capacity \cite{Lapidoth_Moser03}. Moreover, this law also
satisfies the peak power constraint $|X|^2 \leq \Es$, so that the lower
bound on the cut-off rate we compute will also be valid as a lower
bound for the cut-off rate under a peak
constraint. Finally, as the next proposition shows, the fact that
under $\tilde{\law{Q}}$ the input $X$ satisfies, with probability
one, $|X| \geq x_{\textnormal{min}}$, where $x_{\textnormal{min}}
\rightarrow \infty$ greatly simplifies our analysis. It allows us to
asymptotically ignore the additive noise.
\begin{proposition}
  \label{prop:devdeal}
  Let $E_{0}(1, \law{Q}, 0)$ denote the function $E_{0}(\rho, \law{Q},
  r)$ evaluated at $\rho=1, r=0$ for the input law $\law{Q}$ to the
  Ricean channel of specular component $d$ and additive noise variance
  $\sigma^{2}$. Let $E_0^{\sigma =0}(1, \law{Q}, 0)$ be similarly
  defined for the Ricean channel with the same specular component but
  without any additive noise. If under the law $\law{Q}$ the input $X
  \in \Complex$ satisfies with probability one
  \begin{equation*}
    |X| \geq x_{\textnormal{min}}
  \end{equation*}
  for some $x_{\textnormal{min}} > 0$ then 
  \begin{equation}
    \label{eq:simon100}
    E_{0}(1, \law{Q}, 0) \geq E_0^{\sigma =0}(1, \law{Q}, 0) -
    O \left( \frac{\dabs^{2} + 1}{x^{2}_{\textnormal{min}}} \right). 
  \end{equation}
\end{proposition}
\begin{proof}
  For any input probability distribution $\law{Q}$, the term $E_0(1,\law{Q},0)$
  can be expressed
  \begin{align}
     E_0(1,\law{Q},0) & = -\log \int_x \int_{x'}
  \int_y \sqrt{ w(y|x) w(y|x') } \d \mu(y) \d \law{Q}(x') \d
  \law{Q}(x) \nonumber \\
                & = -\log \int_x \int_{x'} B(x,x';\sigma) \d \law{Q}
  (x') \d \law{Q}(x) \label{eq:simon10} 
  \end{align}
  where 
  \begin{equation}
    B(x,x';\sigma) \triangleq \int_y \sqrt{ w(y|x) w(y|x') } \d \mu(y)
  \end{equation}
  and where for the Ricean fading channel with additive noise of variance
  $\sigma^2$
 \begin{equation}
   \label{eq:compB}
   B(x,x';\sigma) =
   \frac{2\sqrt{|x|^2+\sigma^2}\sqrt{|x'|^2+\sigma
       ^2}}{|x'|^2+|x|^2
     + 2\sigma^2}e^{\frac{-\dabs^2 \cdot |x-x'|^2}{2(|x|^2+|x'|^2+2\sigma^2)}}.
 \end{equation}
Comparing $B(x,x';\sigma)$ with the corresponding term in the absence
of noise $B(x,x'; 0)$ we obtain
\begin{IEEEeqnarray}{rCl}
   \IEEEeqnarraymulticol{3}{l}{B(x,x';\sigma) }\nonumber\\\quad
   & \leq & B(x,x';0) \sqrt{1 + \sigma^2 / |x|^2} \sqrt{1
  + \sigma^2 / |x'|^2}e^{\dabs^2 \sigma^2
  \frac{|x-x'|^2}{(|x|^2+|x'|^2+2 \sigma^2)(|x|^2+|x'|^2)}}\\
  & \leq & B(x,x';0) \sqrt{1 + \sigma^2 / |x|^2} \sqrt{1
  + \sigma^2 /
  |x'|^2}e^{|d|^2\sigma^2\frac{(|x|+|x'|)^2}{(|x|^2+|x'|^2+2\sigma^2)(|x|^2+|x'|^2)}}
  \label{eq:simon1} 
\end{IEEEeqnarray}
where the last inequality follows by the triangle inequality. It thus
follows from \eqref{eq:simon10} and \eqref{eq:simon1} that if under
the law $\law{Q}$ the random variable $X$ satisfies with probability
one $|X| \geq x_{\textnormal{min}}$ then 
\begin{align*}
  E_0(1,\law{Q},0) & \geq E_0^{\sigma =0}(1,\law{Q},0) \\ & \quad - \sup_{|x|, |x'| \geq 
  x_{\textnormal{min}}} \Biggl\{ \log\sqrt{1+\sigma^2/|x|^2} 
  +  \log \sqrt{1+\sigma^2/|x'|^2} \\ 
  & \qquad \qquad \qquad + \dabs^2 \sigma^2 \frac{(|x|+|x'|)^2}{(|x|^2+|x'|^2+2\sigma^2)(|x|^2+|x'|^2)}
  \Biggr\} \\
  &  = E_0^{\sigma =0}(1,\law{Q},0) - O \left( 
  (\dabs^{2} + 1)/x^{2}_{\textnormal{min}} \right).
\end{align*}
\end{proof}
Using this proposition with the law $\tilde{\law{Q}}$ under which $X$ is
distributed according to \eqref{eq:simon110} we obtain that 
\begin{equation}
  \label{eq:simon150}
  \varliminf_{\Es \rightarrow \infty} \left\{
  R_{0}(\Es) - E_{0}^{\sigma=0}(1, \tilde{\law{Q}}, 0) \right\}
  \geq 0.
\end{equation}
Computing $E_{0}^{\sigma=0}(1, \tilde{\law{Q}}, 0)$ from
\eqref{eq:simon10} and \eqref{eq:compB} we obtain
\begin{multline}
  \label{eq:zurich1}
  E_{0}^{\sigma=0}(1,\tilde{\law{Q}},0) = -\log 8 + \frac{\dabs^2}{2} +
  2 \log \log \frac{\Es}{\log\Es} \\ - \log
  \int_{\sqrt{\log\Es}}^{\sqrt{\Es}}
  \int_{\sqrt{\log\Es}}^{\sqrt{\Es}} \frac{1}{\rho^2+\rho'^2} \I0
  \left(
    \frac{\dabs^2\rho\rho'}{\rho^2+\rho'^2} \right) \d{\rho} \d{\rho'}.
\end{multline}
The last term on the RHS of the above is difficult to evaluate
precisely. However, since the integrand is positive, the double
integral can be upper bounded by inflating the region of integration
to the region 
\[ \{ \rho, \rho' \geq 0: 2 \log \Es \leq \rho^{2} + \rho'^{2} \leq 2
\Es\}.\]
The integral over this larger set can be now computed analytically by
changing to polar coordinates to obtain
\begin{equation}
  \label{eq:zurich2}
\int_{\sqrt{\log\Es}}^{\sqrt{\Es}}
\int_{\sqrt{\log\Es}}^{\sqrt{\Es}} \frac{1}{\rho^2+\rho'^2} \I0
  \left(
    \frac{\dabs^2\rho\rho'}{\rho^2+\rho'^2} \right) \d{\rho} \d{\rho'}
  \leq \frac{\pi}{2} \I0^2 \left( \frac{\dabs^{2}}{4} \right)
  \cdot \log \sqrt{\frac{\Es}{\log \Es}} 
\end{equation}
where we have used the identity
\begin{equation}
  \label{eq:id_amos}
  \frac{2}{\pi} \int_{0}^{\frac{\pi}{2}} \I0 \bigl( \xi \sin \varphi \bigr)
  \d{\varphi} = \I0^2 (\xi/2), \quad \xi \in \Reals
\end{equation}
which follows from \cite[6.567]{Grad_Ryz_94}.
Consequently, by \eqref{eq:zurich1} and \eqref{eq:zurich2}
\begin{equation}
  \label{eq:E0firm}
  E_{0}^{\sigma=0}(1, \tilde{\law{Q}}, 0) \geq \log \log
  \frac{\Es}{\log \Es} + \frac{\dabs^2}{2} - \log(2 \pi) - 2 \log \I0
        \left(\frac{\dabs^2}{4} \right)
\end{equation}
so that by \eqref{eq:simon150}
\begin{equation}
\label{eq:LoBound}
   \varliminf_{\Es \rightarrow \infty} \left\{ R_0(\Es) - \log \log
        \frac{\Es}{\sigma^2} \right\} \geq 
        \frac{\dabs^2}{2} - \log(2 \pi) - 2 \log \I0
        \left(\frac{\dabs^2}{4} \right).
\end{equation}

\subsection{The Cut-Off Rate in the Presence of Receiver Side Information}

We next consider the case where the fading $H$ is of zero-mean
(Rayleigh) and where the receiver has access to some side-information
$S$ that is jointly Gaussian with $H$. We assume that the pair $(H,S)$
is independent of the additive noise $Z$ and that the joint law of
$(H,S)$ and $Z$ does not depend on the channel input $x \in \Complex$.
We denote the conditional mean of $H$ given $S=s$ by 
\begin{equation}
  \label{eq:def_ds}
  \hat{d}_{s} \triangleq \E{H | S=s}
\end{equation}
and the estimation error by
\begin{equation}
  \label{eq:def_eps2}
  \eps^{2} \triangleq \E{|H - \hat{d}_{s}|^{2} | S=s}.
\end{equation}
Note that unconditionally, $\hat{d}_{s}$ is a zero-mean circularly-symmetric
Gaussian random variable of variance $1- \eps^{2}$:
\begin{equation}
  \label{eq:uncond}
  \hat{d}_{s} \sim {\mathcal N}_{\Complex}(0,1-\eps^{2}).
\end{equation}
Recall also that we only treat here the case $\eps^{2} > 0$.  Denoting the
conditional density of $(Y,S)$ corresponding to the input $x \in
\Complex$ by $w(y,s|x)$, we have by the independence of the side information $S$
and the input that
\begin{equation}
  \label{eq:w_factor}
  w(y,s|x) = f_{S}(s) w(y|x,s)
\end{equation}
where $f_{S}$ is the density of the side information and where $w(y|x,s)$ is the
conditional law of $Y$ given the input $x$ and the side information $s$. Note
that, because $(H,S)$ are jointly Gaussian, the density $w(y|x,s)$ is
the Gaussian density of mean $\hat{d}_{s} \cdot x$ and variance
$\eps^{2} \cdot |x|^{2} + \sigma^{2}$.
Consequently,
\begin{IEEEeqnarray}{rCl}
   \nonumber\\
\IEEEeqnarraymulticol{3}{l}{E_{0}(1, \law{Q}, r) }\nonumber\\\quad
& = & - \log \int_y \int_s 
  \left( \int_x e^{r(|x|^2 - \Es)} \sqrt{w(y,s|x)} \d
  \law{Q}(x)\right)^2 \d \mu(y)\d \mu(s) \nonumber \\
& = & - \log \int_s f_{S}(s) \int_y  
  \left( \int_x e^{r(|x|^2 - \Es)} \sqrt{w(y|x,s)} \d
  \law{Q}(x)\right)^2 \d \mu(y)\d \mu(s) \label{eq:jazz17}
\IEEEeqnarraynumspace \\
& = & - \log \int_{s} f_{S}(s) \cdot \Exp \left( - E_{0}(1, \law{Q}, r | s)
  \right) \d{s} \label{eq:jazz18} 
\end{IEEEeqnarray}
where \eqref{eq:jazz17} follows from \eqref{eq:w_factor} and where
\eqref{eq:jazz18} follows by defining 
\begin{equation}
  \label{eq:def_E0s}
  E_{0}(1, \law{Q}, r | s) \triangleq - \log \int_{y} 
  \left( \int_x e^{r(|x|^2 - \Es)} \sqrt{w(y|x,s)} \d
  \law{Q}(x)\right)^2 \d \mu(y)
\end{equation}
as the $E_{0}$ function corresponding to the channel $w(y|x,s)$
for $S = s$ fixed. (This channel is a Ricean fading channel, except
that the fading is not normalized to have unit variance.)

The cut-off rate $R_{0}(\Es|S)$ in the presence of the side
information $S$ can be thus upper bounded by
\begin{align}
  R_{0}(\Es|S) & \leq \sup_{\law{Q}:\E[\law{Q}]{|X|^{2}} \leq \Es} 
  \sup_{r \geq 0} \left\{
  - \log \int_{s} f_{S}(s) \cdot \Exp \left( - E_{0}(1, \law{Q}, r | s)
  \right) \d{s}\right\} \\
  & \leq - \log \int_{s} f_{S}(s) \cdot \Exp \left( - \sup_{\law{Q}:\E[\law{Q}]{|X|^{2}} \leq \Es} 
  \sup_{r \geq 0} E_{0}(1, \law{Q}, r | s) \right) \d{s} \\
  & = - \log \int_{s} f_{S}(s) \cdot \Exp \left( - R_{0}(\Es|S=s)
  \right) \label{eq:lady1}
\end{align}
where
\begin{equation}
  R_{0}(\Es|S=s) \triangleq \sup_{\law{Q}:\E[\law{Q}]{|X|^{2}} \leq \Es} 
  \sup_{r \geq 0} E_{0}(1, \law{Q}, r | s) \\
\end{equation}
is the  cut-off rate corresponding to power $\Es$
communication over the channel $w(y|x,s)$ for fixed $S=s$.
\footnote{This definition is consistent with (\ref{eq:def_Ezcont})
  since the cost constraint on the cut-off rate is always active for
  the Ricean fading channel.}

It now follows from \eqref{eq:lady1} that
\begin{equation*}
  R_{0}(\Es|S) - \log \log \frac{\Es}{\sigma^2} \leq  - \log \int_{s} f_{S}(s) \cdot \Exp
  \Bigl( - \bigl( R_{0}(\Es|S=s) - \log \log \frac{\Es}{\sigma^2} \bigr) \Bigr)
\end{equation*}
and consequently
\begin{IEEEeqnarray}{rCl}
  \IEEEeqnarraymulticol{3}{l}{
    \varlimsup_{\Es \rightarrow \infty} \{ R_{0}(\Es|S) - \log \log
    \frac{\Es}{\sigma^2}\}}
  \nonumber \\ \qquad
  & \leq & \varlimsup_{\Es \rightarrow \infty} \left\{
  - \log \int_{s} f_{S}(s) \cdot \Exp
  \Bigl( - \bigl( R_{0}(\Es|S=s) - \log \log \frac{\Es}{\sigma^2} \bigr) \Bigr) \d{s}\right\}\\ 
  & = & - \log \varliminf_{\Es \rightarrow \infty} 
  \int_{s} f_{S}(s) \cdot \Exp
  \Bigl( - \bigl( R_{0}(\Es|S=s) - \log \log \frac{\Es}{\sigma^2} \bigr) \Bigr) \d{s}\\
 & \leq & - \log \int_{s} f_{S}(s) \varliminf_{\Es \rightarrow
  \infty} 
\Exp \Bigl( - \bigl( R_{0}(\Es|S=s) - \log \log \frac{\Es}{\sigma^2} \bigr) \Bigr)
  \d{s} \\
& = & - \log \int_{s} f_{S}(s) \cdot
\Exp \Bigl(   - \lim_{\Es \rightarrow \infty} 
\bigl\{ R_{0}(\Es|S=s) - \log \log \frac{\Es}{\sigma^2} \bigr\} \Bigr)
  \d{s} \\
& = & - \log \int_{s} f_{S}(s) \cdot
\Exp  \left( -\frac{|\hat{d}_s|^2}{2\epsilon^2} + \log(2 \pi) + 2 \log \I0
  \left( \frac{|\hat{d}_s|^2}{4\epsilon^2} \right) 
 \right) \d{s} \IEEEeqnarraynumspace \\
& = & \log\frac{1}{\epsilon^2} - 
        \log \K \left( \sqrt{1-\epsilon^4}\right) -\log4.
    \label{eq:jazz56}
\end{IEEEeqnarray}
Here the swapping of the limit and the expectation (second inequality)
is justified using Fatou's lemma and we use the result
\begin{equation}
  \lim_{\Es \rightarrow \infty} \Bigl\{ R_0(\Es| S=s) - \log\log \frac{\Es}{\sigma^2} \Bigr\}
  = \frac{|\hat{d}_s|^2}{2 \eps^2} - \log(2 \pi) - 2 \log \I0
  \left( \frac{|\hat{d}_s|^2}{4\eps^2} \right) 
\end{equation}
which follows from \eqref{eq:cutoff_noSI} applied to the un-normalized
Ricean fading channel whose specular component is $\hat{d}_{s}$ and
whose granular component is of variance $\eps^{2}$. The evaluation of
the last integral is based on an identity combining
\cite[6.612]{Grad_Ryz_94} and \cite[160.02]{Byrd_Fried_71}
\begin{equation*}
  \int_0^{\infty}e^{-\alpha x}(\I0(\beta x))^2\d x =
  \frac{2}{\pi\alpha}K\left(\frac{2\beta}{\alpha}\right), \ \ \alpha,
  \beta >0
\end{equation*}
and the identity for the elliptic function \cite[Eq.\ (3.2.4)]{Andrews_Askey_Roy} 
\begin{equation}
  \label{eq:identity}
  K(k) = \frac{2}{1+k'} K \left( \frac{1-k'}{1+k'} \right),
  \qquad k^{2} + k'^{2} =1, \ 0<k,k'<1.
\end{equation}
In view of \eqref{eq:jazz56}, to establish \eqref{eq:cutoff_with_si}
it now suffices to show
\begin{equation}
  \label{eq:with_si_lim}
  \varliminf_{\Es \rightarrow \infty} \{ R_{0}(\Es|S) - \log \log
  \frac{\Es}{\sigma^2}\} \geq 
\log\frac{1}{\epsilon^2} - 
        \log \K \left( \sqrt{1-\epsilon^4}\right) -\log4.
\end{equation}
To this end we note that by \eqref{eq:jazz18} and \eqref{eq:def_E0s}
evaluated at $r=0$
\begin{equation}
  \label{eq:jazz29}
  R_{0}(\Es|S) \geq - \log \int_{s} f_{S}(s) \cdot \Exp \left( - E_{0}(1, \tilde{\law{Q}}, 0 | s)
  \right) \d{s}
\end{equation}
for any law $\tilde{\law{Q}}$ satisfying $\E[\tilde{\law{Q}}]{|X|^{2}}
\leq \Es$. We next choose, as before, $\tilde{\law{Q}}$ to be a law
under which $X$ is circularly symmetric with
\begin{equation}
  \log |X|^{2} \sim \text{Uniform} 
\left( \log \log \Es, \log \Es \right)
\end{equation}
whence by Proposition~\ref{prop:devdeal} and \eqref{eq:E0firm} applied
to the Ricean channel of fading mean $\hat{d}_{s}$ and granular component
$\eps^{2}$ and the tightness of the lower bound
\begin{equation}
  \label{eq:jazz99}
  \lim_{\Es \rightarrow \infty} \left\{ E_{0}(1, \tilde{\law{Q}}, 0|s) - \log
  \log \frac{\Es}{\sigma^2} \right\} = \frac{|d_{s}|^2}{2 \eps^{2}} - \log(2 \pi) - 2 \log \I0
        \left(\frac{|d_{s}|^2}{4 \eps^{2}} \right)
\end{equation}
for every $s$. The desired result \eqref{eq:with_si_lim} now follows
from \eqref{eq:jazz29} and \eqref{eq:jazz99} using the Dominated
Convergence Theorem and \eqref{eq:uncond}.

\appendix

\section{A Lagrange Duality \label{app:lagrange}}

In this appendix we prove the following Lagrange duality: 
\begin{proposition}
\label{prop:LagDual}
  For any discrete memoryless channel and any $\varrho > 0$, the problem 
  \begin{equation}
   \label{eq:expCsi}
    \min_{\law{Q}}e^{-  \CKEzrhoq(\varrho, \law{Q})}
  \end{equation}
 
is a Lagrange dual of the problem
\begin{equation}
   \label{eq:expGal}
 \min_{\law{Q}}e^{- \GalEzrhoq(\varrho, \law{Q})}
\end{equation}
where $\law{Q}$ is a distribution on the input alphabet.
In particular, since strong duality holds,
\begin{equation*}
   \max_{\law{Q}}\GalEzrhoq(\varrho,\law{Q}) = \max_{\law{Q}} \CKEzrhoq(\varrho,\law{Q})
\end{equation*}
\end{proposition}
\begin{proof}
 Consider a discrete memoryless channel $\law{W}(y|x)$ with input $X \in
 \set{X}$, $|\set{X}|= N$ and output $Y \in \set{Y}$, $|\set{Y}|=M$. We henceforth introduce the more standard, for optimization
problems, vector notation for functions on discrete domains. Hence, let $\bfq
\in \Reals^{1 \times N}$ be a probability distribution on $\set{X}$ and
$\bfw  \in \Reals^{N\times M}$ be a matrix whose (i,j)-th element is
given by
\begin{equation*}
  w_{ij}=\law{W}(y_j|x_i)^\frac{1}{\varrho +1}, \ x_i \in
\set{X}, \ y_j \in \set{Y}, \  \varrho > 0.
\end{equation*}
Hence, (\ref{eq:expGal}) can be written as:
\begin{align*}
& \min_{\bfq,\bfef} \sum_j f_j^{1+\varrho} \\
& \text{s.t.} \ \bfq \bfw = \bfef, \ \bfq \succeq \bfzero, \ \bfq {\bfone} = 1,
\end{align*}
where  $\bfef \in \Reals^{1 \times M}$ is an auxiliary vector that we introduce
in this problem. 
The domain $\const{D}$ of this optimization problem is
$\const{D}=\{(\bfq,\bfef) | \bfq \succeq \bfzero\}$. For any $\varrho
> 0$ the objective function is convex in $\const{D}$.
Furthermore, all equality and inequality constraints are affine.
Hence, the problem is a \emph{convex optimization problem}.
We will perform a relaxation,  which is nevertheless tight for the
optimal values of $\bfef$ and $\bfq$, to the constraint $\bfq \bfw = \bfef$, namely
\begin{align*}
  & \min_{\bfq,\bfef} \sum_j f_j^{1+\varrho} \\
  & \textnormal{s.t.} \ \bfef \succeq \bfq \bfw, \ \bfq \succeq
  \bfzero, \ \bfq {\bfone} = 1.
\end{align*}
The Lagrangian function of this problem is
\begin{equation*}
L(\bfq,\bfef, \bfnu , \mu ,\lam )= \sum_jf_j^{1+\varrho} + (\bfq\bfw - \bfef )\bfnu + (1-\bfq\bf{1})\mu - \bfq \lam,
\end{equation*}
where $\lam \succeq \bfzero \in \Reals^{N \times 1}$, $\bfnu \succeq \bfzero \in
\Reals^{M\times 1}$, $\bfef \in \Reals^{1 \times M}$, $\mu \in
\Reals$ and $(\bfq, \bfef)\in \const{D}$.  Since the Lagrangian
function is affine with respect to $\bfq$, we impose the dual inequality constraint $\mu \bfone
 \preceq \bfw \bfnu $,
 minimize the Lagrangian over $\bfef$ and obtain the Lagrange dual problem 
\begin{align*}
  &\max_{\bfnu, \mu} \left\{-\varrho\sum_j
  \left(\frac{\nu_j}{1+\varrho}\right)^{\frac{1+\varrho}{\varrho}} + \mu\right\} \\
  &\textnormal{s.t.} \ \mu \bf{1} \preceq\bfw \bfnu, \ \bfnu \succeq \bfzero.
\end{align*}
This is a concave problem, with the objective function being monotonic
with respect to all the optimization variables. Since we maximize it in a polyhedron, the optimum will be
on the boundary, of maximum distance from the hyperplane $\mu = 0$ and
of minimum distance from all hyperplanes that define the polyhedron.
Therefore, some dual constraint has to be active, i.e.,
\begin{equation*}
  \min_i
\sum_jw_{ij}\nu_j=\mu.
\end{equation*}
Consequently, the dual problem becomes
\begin{align*}
  &\max_{\bfnu \succeq \bfzero} \left\{- \varrho \sum_j
  \left(\frac{\nu_j}{1+\varrho}\right)^{\frac{1+\varrho}{\varrho}} +
  \min_i\left\{\sum_jw_{ij}\nu_j\right\}\right\}.
\end{align*}
We perform the transformation of variables
$\frac{\nu_j}{1+\varrho}=
r_j^{\frac{\varrho}{1+\varrho}}\alpha,\ j=1, \dots, M$, where
$\vect{r}\in\Reals^{1\times M}$
is chosen to be a probability distribution and $\alpha \in \Reals^+$ is
the appropriate normalizing scalar. Optimizing over $\alpha$ yields
\begin{align*}
  & \max_\vect{r} \left\{\left(\min_i\sum_j w_{ij}r_j^\frac{\varrho}{\varrho
    +1}\right)^{\varrho +1} \right\}\\ & \textnormal{s.t.} \  \vect{r} \succeq \bfzero, \vect{r} {\bfone} = 1
\end{align*}
which, because of the fact that $\left(\sum_j w_{ij}r_j^\frac{\varrho}{\varrho
    +1}\right)^{\varrho +1}$ is concave with respect to $\vect{r}$ and
    monotonic with respect to $\sum_j w_{ij}r_j^\frac{\varrho}{\varrho
    +1}$,
  concludes the proof.
\end{proof}


\section{Proof of \eqref{eq:same_const} \label{app:equal_rc}}

\begin{proof}
  We begin with the case where the cost constraint is active. Fix some
  $\varrho \geq 0$ and let $\law{Q}_{*}$ and $r_{*}$ achieve
  \begin{equation*}
    \max_{\law{Q}: \E[\law{Q}]{g(X)}=\Upsilon} \max_{r \geq 0} E_{0}(\varrho, \law{Q}, r)
  \end{equation*}
  so that
  \begin{equation}
    \label{eq:a1}
    E_{0}(\varrho, \law{Q}_{*}, r_{*}) = 
    \max_{\law{Q}: \E[\law{Q}]{g(X)}=\Upsilon} \max_{r \geq 0} E_{0}(\varrho, \law{Q}, r).
  \end{equation}
  Following \cite[Eq.~(7.3.26)]{Gallager68} we define
  \begin{equation}
    \label{eq:def_alpha}
    \alpha(y) \triangleq \sum_{x \in \set{X}} \law{Q}_{*}(x) e^{r_{*}(g(x)
    - \Upsilon)} \law{W}(y|x)^{\frac{1}{1+\varrho}}, \quad y \in \set{Y}.
  \end{equation}
  With this definition we have by \eqref{eq:a1} and \eqref{eq:modGall}
  \begin{align}
    \max_{\law{Q}: \E[\law{Q}]{g(X)}=\Upsilon} \max_{r \geq 0} E_{0}(\varrho, \law{Q}, r) 
    & = E_{0}(\varrho, \law{Q}_{*}, r_{*}) \nonumber \\
    \label{eq:a17}
    & = - \log \sum_{y \in \set{Y}} \alpha^{1+\varrho}(y).
  \end{align}
Also, by \cite[Eq.~(7.3.28)]{Gallager68} 
\begin{equation}
  \label{eq:kuhn17}
  \sum_{y \in \set{Y}} \alpha^{\varrho}(y) e^{r_{*}(g(x) - \Upsilon)}
  \law{W}(y|x)^{\frac{1}{1+\varrho}}  \geq \sum_{y \in \set{Y}}
  \alpha^{1+\varrho}(y), \quad \forall x \in \set{X}.
\end{equation}
Consider now the distribution $\law{R}_{*}$ on $\set{Y}$ given by
\begin{equation}
  \label{eq:def_Ropt}
  \law{R}_{*}(y) = \frac{\alpha^{1+\varrho}(y)}
  {\sum_{y' \in \set{Y}} \alpha^{1+\varrho}(y')}, \quad y \in \set{Y}.
\end{equation}
We now have by \eqref{eq:amos104} that for any distribution $\law{Q}$ 
\begin{equation}
  \label{eq:tyty12}
  \CKEzrhoq(\varrho, \law{Q}) \leq 
  -(1+\varrho) \sum_{x \in \X} \law{Q}(x) \log \left( \sum_{y \in \Y}
  \law{W}(y|x)^{\frac{1}{1+\varrho}}
    \law{R}_{*}(y)^{\frac{\varrho}{1+\varrho}} \right)
\end{equation}
and if $\E[\law{Q}]{g(X)} = \Upsilon$ then
\begin{IEEEeqnarray*}{rCl}
  \IEEEeqnarraymulticol{3}{l}{\CKEzrhoq(\varrho, \law{Q})
  }\nonumber\\\quad
  & \leq &
  -(1+\varrho) \sum_{x \in \X} \law{Q}(x) \log \left( \sum_{y \in \Y}
  e^{r_{*}( g(x) - \Upsilon )} \law{W}(y|x)^{\frac{1}{1+\varrho}}
    \law{R}_{*}(y)^{\frac{\varrho}{1+\varrho}} \right) \\
  & = & -(1+\varrho) \sum_{x \in \X} \law{Q}(x) \log \left( \sum_{y \in \Y}
  e^{r_{*}( g(x) - \Upsilon )} \law{W}(y|x)^{\frac{1}{1+\varrho}}
    \alpha^{\varrho}(y) \right) + \varrho \log
\sum_{y \in \set{Y}}
  \alpha^{1+\varrho}(y) \\
& \leq & - \log \sum_{y \in \set{Y}} \alpha^{1+\varrho}(y) \\
& = & \max_{\law{Q}: \E[\law{Q}]{g(X)}=\Upsilon} \max_{r \geq 0} E_{0}(\varrho, \law{Q}, r).
\end{IEEEeqnarray*}
Here the first inequality follows from \eqref{eq:tyty12} because the
condition $\E[\law{Q}]{g(X)} = \Upsilon$ guarantees that the
introduction of the exponential term $\textnormal{exp}\{r_{*}(g(x) -
\Upsilon)\}$ has zero net effect; the subsequent equality by
\eqref{eq:def_Ropt}; the subsequent inequality by \eqref{eq:kuhn17};
and the final equality by \eqref{eq:a1}. It thus follows upon taking
the supremum in the above over all laws $\law{Q}$ satisfying
$\E[\law{Q}]{g(X)} = \Upsilon$ that
\begin{equation}
  \label{eq:tyty20}
  \max_{\law{Q}: \E[\law{Q}]{g(X)} = \Upsilon} \CKEzrhoq(\varrho,
  \law{Q}) \leq \max_{\law{Q}: \E[\law{Q}]{g(X)}=\Upsilon} \max_{r \geq 0} E_{0}(\varrho, \law{Q}, r).
\end{equation}
On the other hand, by \eqref{eq:shir3} we obtain
\begin{equation}
  \label{eq:tyty25}
  \max_{\law{Q}: \E[\law{Q}]{g(X)} = \Upsilon} \CKEzrhoq(\varrho,
  \law{Q}) \geq \max_{\law{Q}: \E[\law{Q}]{g(X)}=\Upsilon} \max_{r \geq 0} E_{0}(\varrho, \law{Q}, r)
\end{equation}
which combines with \eqref{eq:tyty20} to prove the claim for active
cost constraints.

For the case of inactive cost constraints we have
\begin{align*}
  \max_{\law{Q}: \E[\law{Q}]{g(X)} \leq \Upsilon} \CKEzrhoq(\varrho,
  \law{Q}) & \leq \max_{\law{Q}} \CKEzrhoq(\varrho, \law{Q})  \\
  & = \max_{\law{Q}} \GalEzrhoq(\varrho, \law{Q}) \\
  & = \max_{\law{Q}: \E[\law{Q}]{g(X)} \leq \Upsilon} 
  \GalErqM(\varrho, \law{Q}).
\end{align*}
Here the first inequality follows by relaxing the constraint; the
subsequent equality by \eqref{eq:jazz30}; and the final equality by
\eqref{eq:max_simple}. This combines with \eqref{eq:shir3} to conclude
the proof.
\end{proof}


\section{Derivation of \eqref{eq:Mim1} \label{app:mim}} 

To derive \eqref{eq:Mim1} we begin by noting that for $\rho \geq 1$ the integrand can be
lower bounded by its value when $\alpha=0$ because
\begin{equation*}
  \left(\rho^2 + \delta\right)^\frac{\alpha-1}{2} \geq  \left(\rho^2 +
  \delta\right)^{-\frac{1}{2}}, \qquad \alpha,\delta \geq 0, \ \rho\geq 1.
\end{equation*}
In the region $0 \leq \rho \leq 1$ we can use the inequality
\begin{equation*}
   \left(\rho^2 + \delta\right)^\frac{\alpha-1}{2} \geq
   \delta^{\frac{\alpha}{2}} \left(\rho^2 +
   \delta\right)^{-\frac{1}{2}}, \ \ \alpha, \delta > 0, \ 0\leq\rho\leq 1.
\end{equation*}
Combining the above two bounds we obtain that throughout the region of
integration
\begin{equation*}
    \left(\rho^2 + \delta\right)^\frac{\alpha-1}{2} \geq
   \delta^{\frac{\alpha}{2}} \left(\rho^2 +
   \delta\right)^{-\frac{1}{2}}, \qquad \alpha > 0, 0 < \delta < 1, \
   0\leq\rho < \infty
\end{equation*}
and hence
\begin{equation}
 \label{eq:lab_G1}
   \ell(x; \alpha, \beta, \delta) \geq 
   \delta^ \frac{\alpha}{2} \cdot \ell(x; \alpha = 0, \beta, \delta).
\end{equation}
 
We next relate $\ell(x; \alpha = 0,\beta,\delta)$ to $\ell(x; \alpha =
0,\beta, \delta = 0)$. To that end denote the integrand in $\ell(x;
\alpha = 0, \beta,\delta)$ by
 \begin{equation*}
   \eta(\rho;x,\beta,\delta,d) = e^{-\rho^2\frac{\beta + |x|^2 +
   \sigma^2}{2\beta(|x|^2 +
   \sigma^2)}}\sqrt{\frac{\rho^2}{\rho^2+\delta}} \I0\left(\frac{\dabs
   \cdot |x| \cdot \rho}{|x|^2+\sigma^2}\right).
 \end{equation*}
We now write the integral as
\begin{equation*}
  \ell(x; \alpha = 0,\beta,\delta)= \int_{0}^{\sqrt{m_1\delta}}
  +  \int_{\sqrt{m_1\delta}}^{\infty} \eta(\rho;x,\beta,\delta,d)\d\rho.
\end{equation*}
In the region $\rho \geq \sqrt{m_1\delta}$ we have
\begin{equation*}
  \sqrt{\frac{\rho^2}{\rho^2+\delta}} \geq \sqrt{\frac{m_1}{m_1+1}}
\end{equation*}
and hence 
\begin{equation}
  \label{eq:zvika1}
  \int_{\sqrt{m_1\delta}}^{\infty}\eta(\rho;x,\beta,\delta,d)\d\rho
  \geq
  \sqrt{\frac{m_1}{m_1+1}} \int_{\sqrt{m_1\delta}}^{\infty}
  \eta(\rho;x,\beta, \delta = 0,d) \d \rho.
\end{equation}

We next show that when $\sqrt{m_1\delta}$ is small, the integral over
the interval $[0,\sqrt{m_1\delta}]$ is also small.
Indeed,
\begin{equation*}
  \frac{|x|}{|x|^2 + \sigma^2} \leq \frac{1}{2\sigma}, \qquad x \in \Complex
\end{equation*}
which combines with the monotonicity of $\I0(\cdot)$ and the
fact that the argument to the exponential function is negative to
demonstrate that
\begin{equation*}
  0 \leq \eta(\rho;x,\beta, \delta = 0, d)\leq \I0 \left(
  \frac{\dabs \cdot \rho}{2\sigma} \right) 
\end{equation*}
and hence that
\begin{equation}
  \label{eq:zvika2}
  0 \leq
  \int_{0}^{\sqrt{m_1 \delta}} \eta(\rho;x,\beta, \delta = 0,d) \d \rho \leq
  \sqrt{m_1\delta} \cdot \I0 \left(
  \frac{\dabs\sqrt{m_1\delta}}{2\sigma} \right).
\end{equation}
On the other hand a straightforward calculation demonstrates that
\begin{align}
  \label{eq:zvika3}
  \int_{0}^{\infty} \eta(\rho;x,\beta, \delta = 0,d) \d \rho & \geq
  \int_0^\infty \eta(\rho;x,\beta, \delta = 0, d = 0) \d \rho 
\nonumber \\
&= \sqrt{\frac{\pi}{2}} \cdot 
  \sqrt{\frac{\beta(|x|^2+\sigma^2)}{\beta+|x|^2+\sigma^2}} 
\nonumber \\
&\geq \sqrt{\frac{\pi \beta \sigma^2}{2(\beta+\sigma^2)}}
\end{align}
where the first inequality follows from the monotonicity of
$\I0(\cdot)$ and the final inequality follows from simple algebra.
We thus conclude that
\begin{IEEEeqnarray}{rCl}
   \nonumber\\
   \IEEEeqnarraymulticol{3}{l}{\ell(x; \alpha = 0, \beta, \delta)
   }\nonumber\\\quad
   & = &
 \int_{0}^{\infty} \eta(\rho;x,\beta, \delta,d) \d \rho  
 \nonumber \\ 
 & \geq & 
\int_{\sqrt{m_1 \delta}}^{\infty} \eta(\rho; x,\beta, \delta,d) \d
 \rho \nonumber \\
 & \geq & \sqrt{\frac{m_1}{m_1+1}} \int_{\sqrt{m_1\delta}}^{\infty}
 \eta(\rho;x,\beta, \delta = 0,d) \d \rho \nonumber \\
& = & \sqrt{\frac{m_1}{m_1+1}} \left( \int_{0}^\infty - 
  \int_0^{\sqrt{m_1\delta}} \eta(\rho;x,\beta, \delta = 0,d) \d \rho
\right) 
\nonumber \\
& = & \sqrt{\frac{m_1}{m_1+1}} \left( 1 - \frac{\int_0^{\sqrt{m_1\delta}} 
\eta(\rho;x,\beta, \delta = 0,d)\d \rho}
{\int_0^\infty \eta(\rho;x,\beta, \delta = 0,d) \d \rho} \right)
 \nonumber \\
& & \qquad \qquad \qquad \qquad \qquad \qquad \qquad \cdot 
\int_0^\infty \eta(\rho;x,\beta, \delta = 0,d) \d \rho 
\nonumber \\
&\geq & \sqrt{\frac{m_1}{m_1+1}} \left( 1 -
  \frac{\sqrt{m_1\delta} \cdot \I0
    \left(\frac{\dabs\sqrt{m_1\delta}}{2\sigma}\right)}
{\sqrt{\frac{ \pi \beta\sigma^2}{2(\beta+\sigma^2)}}}\right) 
\ell(x; \alpha=0, \beta, \delta = 0)\label{eq:lab_G2}
\end{IEEEeqnarray}
where the first inequality follows from the non-negativity of the
integrand; the subsequent inequality from \eqref{eq:zvika1}; and the
final inequality from \eqref{eq:zvika2} \& \eqref{eq:zvika3}. The
desired bound \eqref{eq:Mim1} now follows from \eqref{eq:lab_G2} and
\eqref{eq:lab_G1}.


\end{document}